\documentclass[12pt,preprint]{aastex}

\usepackage{wrapfig}

\def\lapprox{\mathrel{\hbox{\rlap{\hbox{\lower4pt\hbox{$\sim$}}}\hbox{$<$}}}}
\def\gapprox{\mathrel{\hbox{\rlap{\hbox{\lower4pt\hbox{$\sim$}}}\hbox{$>$}}}}

\newcommand{\be}{\begin{equation}}
\newcommand{\ee}{\end{equation}}

\begin{document}

\shorttitle{Ion-Neutral Coupling and Flux Emergence}
\shortauthors{Leake et al.}

\title{Effect of ion-neutral collisions in simulations of emerging active regions}

\author{James E. Leake}
\affil{College of Science, George Mason University, 4400 University Drive, Fairfax, Virginia 22030. \\
jleake@gmu.edu}

\and 
\author{Mark G. Linton}
\affil{U.S. Naval Research Lab 4555 Overlook Ave., SW Washington, DC 20375.}
\begin{abstract}

We present results of 2.5D numerical simulations of the emergence of sub-surface magnetic flux into the solar atmosphere, with emerging flux regions ranging from $10^{18}$ to $10^{21}$ Mx, representing both ephemeral and active regions. We include the presence of neutral Hydrogen in the governing equations, improve upon previous models by including the ionization in the equation of state, and use a more realistic convection zone model. We find that ionization and recombination of plasma during the rise of a convection zone flux tube reduces the rise speed of the tube's axis. 
The presence of neutral Hydrogen allows the effective flow of mass across fieldlines, by the addition of a Pedersen resistivity to the generalized Ohm's law, which dissipates current perpendicular to the magnetic field. This causes an increase of up to 10\% in the amount of magnetic in-plane flux supplied to the corona and a reduction of up to 89\% in the amount of sub-surface plasma brought up into the corona. However, it also reduces the amount of free magnetic energy supplied to the corona, and thus does not positively affect the likelihood of creating  unstable coronal structures.

\end{abstract}

\section{INTRODUCTION}

Coronal mass ejections (CMEs) and flares are the most energetic manifestations of solar activity. 
It is generally accepted that the energy required for these events is stored as magnetic 
energy in the corona. This free magnetic energy is most likely stored in
 the strongly
sheared flux of filament channels (see, e.g., reviews by
\citet{2000JGR...10523153F,Klimchuk2001,LintonM2009}).  
These strongly sheared structures have a significant component of the magnetic field  
 parallel to the neutral line, compared to a potential configuration where the field is perpendicular 
 to the neutral line.
In the strong complex regions that are sources of fast CMEs, these 
 filament channels form with the active region \citep{1995JGR...100.3355F} and 
hence the shear field must emerge with the flux. Given that shear
formation is a fundamental driver of the eruption in CME models such as the magnetic breakout model \citep{1999ApJ...510..485A} and the flux cancellation model \citep{2000ApJ...529L..49A}, it is clear that
flux emergence must be explicitly included in such CME models
for them to be physically rigorous.

Most CME models, such as those of \citet{1999ApJ...510..485A}, \citet{2000ApJ...529L..49A}, \citet{2000ApJ...545..524C}, and \citet{2007ApJ...668.1232F} are constrained by the need to extend the simulation domain to at least a few solar radii. They therefore do not model the lower solar atmosphere. 
 The lower boundary of these simulations has a typical density 
of $3\times10^{-16} \ \textrm{g/cm}^3$ and a typical plasma-$\beta<0.2$, where
\be
\beta = \frac{\mu_{0} P}{B^{2}},
\ee
$P$ is the gas pressure, $B$ is the magnetic field strength, and $\mu_{0}$ is the permeability of free space.
The photosphere, in contrast,  has a
 density of around 
$3\times10^{-7} \ \textrm{g/cm}^{3}$ \citep{1981ApJS...45..635V} and a $\beta>1$ \citep{1999SoPh..186..123G,2001SoPh..203...71G}.  The supply of magnetic free energy to the corona in active regions is due to the emergence of sub-surface magnetic field structures, and so 
simulations of CME initiation must therefore address the critical question of whether and how newly emerging, sheared magnetic flux can rise from its origins in the high $\beta$ convection zone to the low corona where it is required to drive many CME models.  

The problem of flux emergence has been studied extensively for a few decades. The difficulty in studying flux emergence is that a simulation must couple the high $\beta$, high density convection zone, where plasma pressure forces dominate, and the corona, which is low $\beta$ and magnetically dominated. The basic idea of flux emergence is that  sub-surface flux tubes are created by dynamo actions at the base of the convection zone. Indeed recent studies suggest that deep convective layers control the pattern of large-scale solar activity \citep{2011SoPh..270....1A,2012SoPh..tmp..244A}. These flux tubes are assumed to have acquired sufficient twist to survive the convective motions as they rise to the surface.
The subsequent expansion into the atmosphere and its interaction with both a field-free corona and pre-existing coronal fields has been studied extensively in recent years, and a comprehensive list can be found in the review of \citet{2008JGRA..11303S04A}.

In \citet{2010ApJ...722..550L} we showed that emerging two and a half dimensional (2.5D) sub-surface flux tubes did not supply enough free energy to the corona (less than 20\% of the magnetic energy was in the shear component of the field) and hence the resulting coronal magnetic field did not erupt.  We concluded that in order to create an eruption a method is needed which allows the transfer of significant shear field into the corona by removing the dense material which is inhibiting the rise of the emerging structures.
This can be achieved in two ways. One method is the draining of plasma along the axis of the emerging flux tube. Recent three dimensional (3D) simulations of flux emergence have shown that the drainage along strongly arched flux tube axes can aid the rise of magnetic field into the corona \citep{2009A&A...503..999H,2009A&A...507..995M}. Also, complex 3D motions associated with shear flows during flux emergence have helped increase the amount of free energy that emerges into the corona \citep{2004ApJ...610..588M,2008A&A...492L..35A,2009A&A...508..445M}. A second method is the drainage of neutral plasma across field lines. 
Using a model for the support of a prominence in a partially ionized solar atmosphere, \citet{2002ApJ...577..464G}  predicted the amount of vertical  draining of neutral atoms from prominence structures, and also found evidence of cross-field diffusion of neutral material in solar filaments to support this model \citep{2007ApJ...671..978G}. In this paper we will explore the drainage of neutral material across fieldlines during flux emergence, using a modified MHD model which includes the partial ionization of the solar atmosphere.

In the simulations of \citet{2010ApJ...722..550L} and the majority of previous simulations, the numerical models use the magnetohydrodynamics (MHD) equations for a fully ionized plasma.  For most of the convection zone and corona this is a valid approach. However, in the upper 3 Mm of the convection zone, as well as  the photosphere and the chromosphere,	 the temperature is low enough that the plasma is not fully ionized. It has been shown that the effects of ion-neutral collisions create an anisotropic resistivity in the single-fluid equations \citep{Cowling1957,1965RvPP....1..205B}. This \textit{Pedersen resistivity} acts only on currents perpendicular to the magnetic field, and has been shown to be up to 12 orders of magnitude larger than the parallel Spitzer, or Coulomb, resistivity in the chromosphere \citep{2004A&A...422.1073K}. The anisotropic dissipation of perpendicular currents by this Pedersen resistivity has been included in the study of the damping of MHD waves in the chromosphere \citep{1999A&A...347..696D, 2005A&A...442.1091L}, and flux emergence in both 2.5D and 3D simulations \citep{2006A&A...450..805L,2007ApJ...666..541A}. In those flux emergence simulations, it was shown that the Pedersen resistivity led to increased rates of flux emergence due to dissipation by ion-neutral collisions, and increased collisional heating.  

In the simulations presented here we improve on two of the limitations of the partially ionized flux emergence simulations of \cite{2006A&A...450..805L} and \citet{2007ApJ...666..541A}. The first is related to the equation of state used. In the simulations of \citet{2006A&A...450..805L} and \citet{2007ApJ...666..541A}, the ionization fraction was calculated as a function of density and temperature, but the specific internal energy density ($\epsilon$) did not include a contributing term from the ionization fraction: The equation of state in those simulations did not include the change 
in internal energy density due to ionization and recombination.  In the simulations presented here, this term is specifically included in the equation of state. As will be shown in the following section, this approach therefore also includes a correct description of a partially ionized convection zone which is adiabatically stratified. 

The second improvement is to use flux tubes which initially contain axial magnetic flux comparable to observations of real active regions on the Sun. The simulations of \citet{2006A&A...450..805L} and \citet{2007ApJ...666..541A} use flux tubes in the convection zone with axial fluxes of less than $10^{19}$  Mx. The axial flux in these tubes is two orders of magnitude less than that typically measured in active regions on the surface of the Sun ($10^{21}$ Mx). More recently, \citet{2010ApJ...720..233C}, \citet{2011ApJ...740...15R}, and \citet{2012ApJ...745...37F} have simulated the emergence of active region size flux tubes ($ > 10^{21}$ Mx). However, those simulations do not include the effects of Pedersen resistivity on the flux emergence process. We model the emergence of large-scale tubes with axial fluxes of $10^{21} $ Mx, as well as the more typical small scale flux tubes seen in the simulations of \citet{2006A&A...450..805L}, \citet{2007ApJ...666..541A}, and \citet{2010ApJ...722..550L}, using our partially ionized plasma model.

\section{NUMERICAL METHOD}
\subsection{Equations}
We solve the standard resistive MHD equations, modified to include the effects of partial ionization. These are solved numerically using the Lagrangian 
remap code \textit{Lare2d} \citep{2001JCoPh.171..151A}. The equations are obtained by summing the equations for ions ($i$) electrons ($e$) and neutrals ($n$).  Hence the total mass density ($\rho$), gas pressure ($P$) and internal specific energy density ($\epsilon$) are obtained by summing over the three species, e.g., $\rho = \sum_{a}{\rho_{a}} = \sum_{a}{m_{a}n_{a}}$ where $m_{a}$ and $n_{a}$ are the mass and number density, respectively, of each species ($a=i,e,n$).  The average velocity, $\mathbf{v}$, is defined by
 $\rho \mathbf{v} = \sum_{a}{\rho_{a}\mathbf{v}_{a}}$.
 The neutral fraction is defined by
\begin{equation}
\xi_{n} = n_{n}/(n_{n}+n_{i}).
\end{equation} 
The equations are
given below in Lagrangian form, using SI units:
\begin{eqnarray}
\frac{D\rho}{Dt} & = & -\rho\nabla.\mathbf{v}, \\
\frac{D\mathbf{v}}{Dt} & = & -\frac{1}{\rho}\nabla P 
+ \frac{1}{\rho}\mathbf{j}\wedge\mathbf{B} + \mathbf{g} + \frac{1}{\rho}\nabla.\mathcal{S},\\
\frac{D\mathbf{B}}{Dt} & = & (\mathbf{B}.\nabla)\mathbf{v} 
- \mathbf{B}(\nabla .\mathbf{v}) - \nabla \wedge (\eta\mathbf{j}_{\|})
-\nabla\wedge(\eta_{p}\mathbf{j}_{\bot}), ~ \textrm{and}\\
\frac{D\epsilon}{Dt} & = & -\frac{P}{\rho}\nabla .\mathbf{v}
 + \varsigma_{ij}S_{ij} + \eta {j_{\|}}^{2} + \eta_{p}{j_{\bot}}^{2} , 
\label{eqn:energy_MHD}
\end{eqnarray}
where $\mathbf{j}_{\|}$ and $\mathbf{j}_{\bot}$ are the parallel and perpendicular current vectors, respectively, and are defined as
\begin{eqnarray}
\mathbf{j}_{\|} & = & \frac{(\mathbf{B}.\mathbf{j})\mathbf{B}}{{|\mathbf{B}|}^{2}},\\
\mathbf{j}_{\bot} & = & \frac{\mathbf{B}\wedge(\mathbf{j}\wedge\mathbf{B})}{{|\mathbf{B}|}^{2}}.
\end{eqnarray}
Here the current density is defined by  $\mathbf{j}=\nabla\wedge\mathbf{B}/\mu_{0}$ and $\mu_{0}$ is the permeability of free space. The total gas density, pressure, and internal specific energy density are 
defined at the center of each numerical cell.
The magnetic field, $\mathbf{B}$,  is defined at cell faces, and the 
 velocity is defined at cell vertices. This staggered grid preserves 
$\nabla.\mathbf{B}$ during the simulation. The gravitational acceleration is denoted by $\mathbf{g}$ and is set to the value of gravity at the mean solar surface ($274 ~ \textrm{m}/\textrm{s}^{2}$). $\mathcal{S}$ is the stress tensor which has components 
$\mathcal{S}_{ij}=\nu(\varsigma_{ij}-\frac{1}{3}\delta_{ij}\nabla.\mathbf{v})$, with
$\varsigma_{ij}=\frac{1}{2}(\frac{\partial v_{i}}{\partial x_{j}}+
\frac{\partial v_{j}}{\partial x_{i}}).$ The viscosity $\nu$ is set to $3\times10^{3} ~ \textrm{kg}~\textrm{m}^{-1}\textrm{s}^{-1}$, and $\delta_{ij}$ is the Kronecker delta function.
The Coulomb resistivity, $\eta$, is given by 
\begin{equation}
\eta = \frac{m_{e}(\nu^{'}_{ei}+\nu^{'}_{en})}{n_{e}e^2},
\end{equation}
where $m_{e}$ and $e$ are the mass and charge of the electron, respectively. The effective collisional frequencies for collisions of electrons with ions and neutrals, respectively, are $\nu^{'}_{ei}$ and
$\nu^{'}_{en}$. The Pedersen resistivity, 
 $\eta_{p}$, is given by
\begin{equation}
\eta_{p} = \eta + \frac{\xi_{n}^{2}{|\mathbf{B}|}^{2}}{\alpha_{n}}.
\label{eqn:etac}
\end{equation}
 The quantity $\alpha_{n}$ is calculated using $\alpha_{n} = m_{e}n_{e}\nu^{'}_{en} + m_{i}n_{i}\nu^{'}_{in}$ where $\nu^{'}_{in}$ is the effective collisional frequency for collisions of ions with neutrals.
We refer the reader to \citet{2005A&A...442.1091L} for the method of calculating the effective collisional frequencies.

For this partially ionized plasma, the total pressure, $P$, and the specific internal energy density, $\epsilon$, can be written as
\begin{eqnarray}
P & = & \rho k_{B}T/\mu_{m}, ~ \textrm{and} \\
\epsilon & =  & \frac{k_{B}T}{\mu_{m}(\gamma-1)} + (1-\xi_{n})\frac{X_{i}}{m_{i}},  
\label{eqn:eos}
\end{eqnarray}
respectively,
where $k_{B}$ is Boltzmann's constant, $\gamma$ is 5/3, and $X_{i}=13.6$ eV is the first ionization energy of Hydrogen. 
Previous simulations of partially ionized flux emergence \citep{2006A&A...450..805L,2007ApJ...666..541A} did not include the 
second term in Equation (\ref{eqn:eos}). In this new model we allow for changes in energy to change the temperature and the ionization level simultaneously.
The neutral fraction, $\xi_{n}$, is a function of temperature, $T$, itself and so Equation (\ref{eqn:eos}) is solved implicitly for $T$ at each time step. The reduced mass, $\mu_{m}$, is 
\begin{equation}
\mu_{m}=m_{i}/(2-\xi_{n}).
\label{eqn:mu}
\end{equation}
Here $m_{i} = m_{f}m_{p}$, where $m_{p}$ is the mass of a proton, and $m_{f}=1.25$ is a pre-factor which is designed to include the effects of heavier elements and, as will be shown in the next section, will help reconcile our initial conditions with more realistic theoretical models of the solar convection zone. 

To calculate $\xi_{n}$ we use a simple model based on the modified Saha equation \citep{1921RSPSA..99..135S}, which takes into account the fact that the chromosphere is not in local thermodynamic equilibrium \citep{1973SoPh...29..421B}.
This equation can be solved for the  steady state solution of the ionization equation \citep{1961psc..book.....A} to give 
\begin{eqnarray}
\label{eqn:saha}
\frac{{n_{i}}^2}{n_{n}} & = & \frac{f(T)}{b(T)}, \\
~\textrm{where} & & \\ \nonumber
f(T) & = & \frac{(2\pi
  m_{e}k_{B}T)^{\frac{1}{\gamma -1}}}{h^{3}}\exp({-\frac{X_{i}}{k_{B}T}}) \label{ion1} \\
  \textrm{and} & & \\ \nonumber
b(T) & = & \frac{T}{wT_{R}}\exp \left[\frac{X_{i}}{4k_{B}T}(\frac{T}{T_{R}}-1
) \right].
\end{eqnarray}
Here $T_{R}$ is the temperature of the photospheric radiation field, $w$ is its dilution factor, and $h$ is the Planck constant. Below the surface,
$T_R=T$ and $w=1$ so that $b(T)=1$. Above the surface, $T_{R}=6420 ~ \textrm{K}$ and $w=0.5$.
Given $n_{i}^{2}/n_{n}$, the ratio of the number density of neutrals to number density of ions
can be calculated from
\begin{equation}
r = \frac{n_{n}}{n_{i}} = 
\frac{1}{2}\left(-1+\sqrt{\left(1+\frac{4\rho/m_{i}}{n_{i}^2/n_{n}}\right)}\right),
\label{ion2}  
\end{equation}
and the neutral fraction, $\xi_{n} = n_{n}/(n_{n}+n_{i})$, is 
\begin{equation}
\xi_{n} = \frac{r}{1+r}.
\label{ion3}
\end{equation}

For the convection zone and lower solar atmosphere, the fully ionized value of the Spitzer/Coulomb resistivity gives a magnetic Reynold's number $R_{m}=UL4\pi/c^{2}\eta$ (where $U$ and $L$ are typical measures of velocity and length) much larger than unity ($ >10^{5}$). Thus typical numerical diffusion in MHD simulations of flux emergence exceeds the classical diffusion, and hence $\eta=0$ is often assumed. On the other hand \citet{2006A&A...450..805L} showed that the value of the Pedersen resistivity gives an effective magnetic Reynold's number below unity in the chromosphere for flux emergence simulations, and that the increased diffusion due to the Pedersen resistivity is significantly larger than both the diffusion due to the Coulomb resistivity and typical numerical diffusion. 

The equations are solved in 2.5D: the simulation box is 2D, with $x$ and $y$ being 
independent variables and $z$ being  ignorable, but all three components of the vector variables are evolved.

\subsection{Background Stratification} 

We require a background stratification which represents the solar convection zone and atmosphere. 
We first define a temperature profile  which is chosen to resemble the real Sun. Then with a specific value of density (or pressure) at the surface, the hydrostatic equilibrium equation can be solved to give the density and pressure in the simulation domain.

Above the surface ($y > 0$), a simple temperature profile is used
\be
T(y) = T_{ph} + \frac{(T_{cor}-T_{ph})}{2}(\tanh{\frac{y-y_{cor}}{w_{tr}}}+1),
\label{eqn:Temp}
\ee
where $T_{ph} = 5778 ~ \textrm{K}$,  $T_{cor} = 150 ~ T_{ph}$, $y_{cor} = 3.75 ~ \textrm{Mm}$ and $w_{tr} = 0.75 ~ \textrm{Mm}$.
This profile is designed to resemble semi-empirical atmospheric models of the quiet Sun developed by \cite{1981ApJS...45..635V}.
Below the surface, an adiabatically stratified gas in hydrostatic equilibrium is used as a model for the atmosphere:  The vertical ($y$) temperature gradient is equal to the adiabatic temperature gradient \citep{STIX}. In this way the atmosphere is marginally stable to convection. For a fully ionized plasma this yields a temperature profile in the convection zone of 
\be
\left(\frac{dT}{dy}\right)_{a} = -\frac{\mu_{m}g}{k_{B}}\frac{\gamma -1}{\gamma}.
\label{eqn:ad_temp_FIP}
\ee

Previous simulations of flux emergence which assume the plasma to be fully ionized typically use the neutral limit of the reduced mass of $\mu_{m}=m_{i}$, rather than $\mu_{m}=m_{i}/2$, which is strictly the fully ionized limit. As we will show, this choice is made to ensure realistic coronal densities result from the solution to the hydrostatic equilibrium equation.
 
Removing the assumption that the plasma is fully ionized yields a different result for the initial equilibrium. Rising and falling packets of plasma can now ionize and recombine, and do not expand and contract ideally.

We want to derive an equation for $ \left(\frac{dT}{dy}\right)_{a}$, the adiabatic temperature gradient for a partially ionized plasma.
This can be done by considering the adiabatic change of a volume $V$ of partially ionized plasma.
Using Equations (\ref{eqn:eos}) and (\ref{eqn:mu}), and defining $n=n_{i}+n_{n}$, and $\zeta=n_{i}/n=1-\xi_{n}$, we can express the energy in volume $V$
as
\be
E = \rho\epsilon V = m_{i}nV\left( \frac{k_{B}T(1+\zeta)}{(\gamma-1)m_{i}} + \frac{\zeta X_{i}}{m_{i}} \right) = N\left( \frac{k_{B}T(1+\zeta)}{(\gamma-1)}+ \zeta X_{i} \right),
\label{eqn:E}
\ee
where $N=nV$ is the total number of ions and neutrals and does not change.

Adiabatic evolution assumes no heat exchange: $dQ=0$. Using the first law of thermodynamics ($dQ = dE + PdV$), this implies
\be
dE = -PdV.
\label{eqn21}
\ee
Differentiating Equation (\ref{eqn:E}) gives
\be
dE = \frac{1}{\gamma -1}Nk_{B}dT(1+\zeta) +\frac{1}{\gamma -1}Nk_{B}Td\zeta +  NX_{i}d\zeta.
\label{eqn23}
\ee
Writing the total pressure as $P = \frac{Nk_{B}T(1+\zeta)}{V}$ we can differentiate $V$ to obtain 
\be
-\frac{dV}{V} = -\left(\frac{-dP}{P} + \frac{dT}{T} + \frac{d\zeta}{1+\zeta}\right).
\label{eqn24}
\ee
 Inserting Equations (\ref{eqn23}) and (\ref{eqn24}) into Equation (\ref{eqn21}), and using $P = \frac{Nk_{B}T(1+\zeta)}{V}$ gives
 \be
\frac{dP}{P} = \frac{\gamma}{\gamma-1}\frac{dT}{T} + \theta\frac{d\zeta}{1+\zeta},
\label{eqn:dta1}
 \ee
 where 
 \be
 \theta \equiv \frac{\gamma}{\gamma-1} + \frac{X_{i}}{k_{B}T}.
 \ee
 This is the condition for the adiabatic change for a partially ionized plasma, with arbitrary ionization state $\zeta$. For practical purposes we would like to eliminate $d\zeta$ from the equation, which we can do by rewriting the Saha equation (Eq. \ref{eqn:saha}) for $y<0$ in terms of $\zeta$:
\be
\frac{\zeta^{2}}{1-\zeta^{2}} = A \frac{T^{\frac{\gamma}{\gamma -1}}}{P}e^{-X_{i}/k_{B}T},
\label{eqn:saha2}
\ee
where $A$ contains physical constants only.
This can be differentiated to give
\be
\frac{2\zeta}{(1-\zeta^{2})^{2}}d\zeta = \frac{\zeta^{2}}{1-\zeta^{2}}\left(-\frac{dP}{P} + \theta\frac{dT}{T}\right),
\ee
or
\be
\frac{d\zeta}{1+\zeta} = \left(-\frac{dP}{P} + \theta\frac{dT}{T}\right)\frac{\zeta(1-\zeta)}{2},
\ee
which can be inserted into Equation (\ref{eqn:dta1})
to give 
\be
\frac{dT}{T}  = \frac{dP}{P}\left(\frac{ 1 + \theta\frac{\zeta(1-\zeta)}{2} }{\frac{\gamma}{\gamma-1}+ \theta^{2}\frac{\zeta(1-\zeta)}{2} }\right).
\label{eqn:dta2}
\ee
Dividing by $dy$, and using  $\frac{dP}{dy} = -\rho g$ and $P=\rho k_{B} T/\mu_{m}$, gives
\be
\left(\frac{dT}{dy}\right)_{a} = -\frac{\mu_{m}g}{k_{B}}\left(\frac{ 1 + \theta\frac{\zeta(1-\zeta)}{2} }{\frac{\gamma}{\gamma-1}+ \theta^{2}\frac{\zeta(1-\zeta)}{2} }\right).
\label{eqn:dta3}
\ee
If $\zeta=0$ or $\zeta=1$,  then we recover the usual definition for a fully neutral or fully ionized plasma, as expressed in Equation  (\ref{eqn:ad_temp_FIP}).

In summary, to construct a model atmosphere which is in hydrostatic equilibrium and whose temperature gradient below the photosphere ($y=0$) matches Equation (\ref{eqn:dta3}), we must perform the following procedure:

\begin{enumerate}
	\item{For $y\ge0$ use the analytic profile $T(y)$ from Equation (\ref{eqn:Temp}).}
	\item{For the entire domain, initially set $\zeta=1$. Set $T(y=0)=5778 ~ \textrm{K}$ and $\rho(y=0)=3.03\times10^{-4} ~ \textrm{kg}/\textrm{m}^3$, taken from the standard solar model presented in \cite{STIX}.}
	\item{For $y<0$, numerically solve Equation (\ref{eqn:dta3}) with boundary condition $T(y=0)$, using the current $\zeta(y)$. This gives $T(y<0)$.}
	\item{For the whole domain, numerically solve the equation $\frac{dP}{dy} = -\rho g$ with boundary condition $\rho(y=0)$, using the current $T(y)$ and $\zeta(y)$, and the equation $P=\rho k_{B} T/\mu_{m}$. This gives $\rho(y)$ for the whole domain.}
	\item{Use the Saha equation (\ref{eqn:saha}) with the current $T(y)$ and $\rho(y)$ to get $\zeta(y)$, and hence the reduced mass $\mu_{m}(y)=\frac{m_{i}}{1+\zeta(y)}$ for the whole domain.}
	\item{Repeat steps 3-5 until the reduced mass $\mu_m(y)$ converges for all $y$.}	
\end{enumerate}

\begin{figure}
\begin{center}
\includegraphics[width=\textwidth]{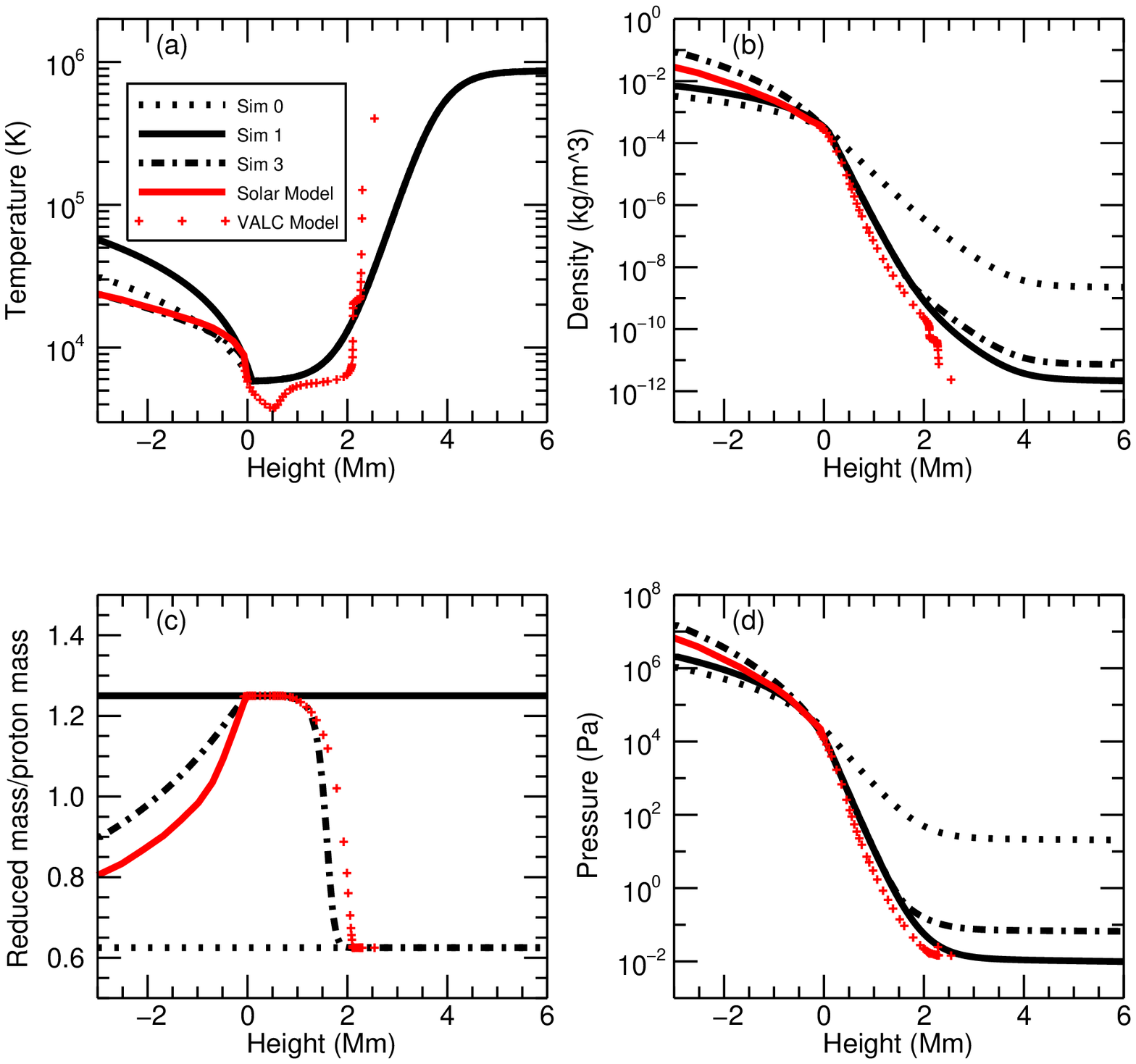}
\caption{Background 1D stratification for three models (Simulations 0, 1, 3).  Dotted line: Simulation 0, the fully ionized simulation with $\mu_{m}=m_{i}/2$.  Solid line: Simulation 1, the fully ionized simulation with $\mu_{m}=m_{i}$. Dot-dashed line: Simulation 3, the partially ionized model. The red line is from a standard model of the solar convection zone \citep{STIX}. The red crosses are the 1D semi-empirical atmospheric model of \citet{1981ApJS...45..635V}. Panel (a): Temperature. Panel (b): Gas density. Panel (c): Reduced mass normalized by the proton mass. Panel (d): Gas pressure.
\label{fig:IC}}
\end{center}
\end{figure}

In order to show how this approach differs from the standard approach used in previous fully ionized simulations, we present the results of this process for three models. A summary of these simulations, along with simulations described in later sections, is shown in Table 1. Simulation 0 uses the  fully ionized plasma model with the ionized limit of $\mu_{m}=m_{i}/2$. Simulation 1 again uses the  fully ionized plasma model  but with the neutral limit of $\mu_{m}=m_{i}$. This is the case used by previous fully ionized simulations \citep{2008JGRA..11303S04A}. We describe Simulation 2 later, and do not consider it at this point. Simulation 3 uses the partially ionized plasma model with an ionization which is allowed to change in space and time, based on the modified Saha equation. The results are shown in Figure \ref{fig:IC}. Also shown are curves taken from a standard solar model which includes the transfer of heat by convection and radiation in the partially ionized plasma of the solar convection zone  \citep{STIX}, and a 1D semi-empirical atmospheric model by \citet{1981ApJS...45..635V}. We can see from Figure \ref{fig:IC} that the temperature of the partially ionized model (Simulation 3) best matches the standard solar model in the convection zone. The temperature of the two fully ionized models (solid and dotted lines) depart from the standard solar model curve in the convection zone, as the ionization cannot change in these two models.
In all of the models we use $m_{f}=1.25$ in order to include the effect of heavier elements, and this ensures that at a height of $y=0$ the models have a reduced mass of $1.25m_{p}$ which is the same as the standard solar model. However, only the partially ionized model is able to capture the variation  in reduced mass with height in the convection zone and corona.
 
While the temperature in the convection zone in the partially ionized model (Simulation 3) matches the standard solar model quite closely, there is some difference between the density and pressure profiles in the convection zone for the partially ionized model and the standard solar model. We expect that this difference is because the models here do not include the ionization of Helium, which is taken into account in the standard solar model. To test whether this difference was caused by the fact that the convection zone is not adiabatically stable throughout its depth, we also tested a model which had $dT/dy = F(dT/dy)_{a}$ where $F=F(y)$ was a function which reproduced the non-adiabatic temperature profile of the standard solar model of \citet{STIX}. This unstable convection zone profile did not produce any significant improvements in the density and pressure profiles. 


Looking at the density and pressure at and above 2 Mm in Figure \ref{fig:IC}, there is a large difference between the fully ionized model with the ionized limit of $\mu_{m}=m_{i}/2$ (Simulation 0) and the other two models, Simulations 1 and 3.  The atmospheric models of \cite{1981ApJS...45..635V} give a density in the upper chromosphere/low corona (at a height of 2 Mm) of $10^{-11} ~ \textrm{kg}/\textrm{m}^{3}$, whereas Simulation 0 has a density 4 orders of magnitude  higher than this. Simulation 0 was included here to show why previous authors who have performed fully ionized simulations use a reduced mass of  $\mu_{m}=m_{i}$ rather than $\mu_{m}=m_{i}/2$. Using the fully ionized limit of $\mu_{m}=m_{i}/2$ in a fully ionized simulation, which would seem the correct choice, actually creates large deviations from coronal densities.
From now on Simulation 0 is not used in this paper.

\section{EMERGENCE OF SMALL SCALE FLUX TUBES}

\subsection{Initial conditions}

In order to compare our partially ionized simulations to previous small-scale flux emergence studies \citep{2001ApJ...549..608M,2001ApJ...554L.111F,2004ApJ...610..588M}, we choose an initial flux tube with a very small axial flux compared to the flux of a real active region. In the next section we will address the emergence of flux tubes with active region size flux ($10^{21}$ Mx).
We insert a cylindrical magnetic flux tube into the model convection zone at $x=0$ and
$y=y_{t}=-1.8 \ \textrm{Mm}$. The axial field, $B_{z}$, and twist field, $B_{\theta}$,  for the tube are given by
\begin{eqnarray}
B_{z} & = & B_{0} {e}^{-r^{2}/a^{2}}, \textrm{and} \\
B_{\theta} & = & qrB_{z},
\end{eqnarray}
respectively,
where $r=\sqrt{x^{2}+(y-y_{t})^{2}}$ is the radial distance 
from the center of the tube, $B_{0}$ is the axial field strength at the center ($r=0$),
$a=0.3 \ \textrm{Mm}$ is the radius of the tube, and the twist $q=-1/a$. For a field strength of  $B_{0}=6000 ~\textrm{G}$, this tube has an axial flux of $5\times10^{18}$ Mx. This tube contains 25\% less axial flux than our previous 2.5D simulations \citep{2010ApJ...722..550L}. 
Our simulation domain is -3 to 42 Mm in $y$ and -22.5 to 22.5 Mm in $x$. 
The numerical grid contains $640\times640$ cells, at a uniform resolution of 70 km.

\begin{figure}
\begin{center}
\includegraphics[width=\textwidth]{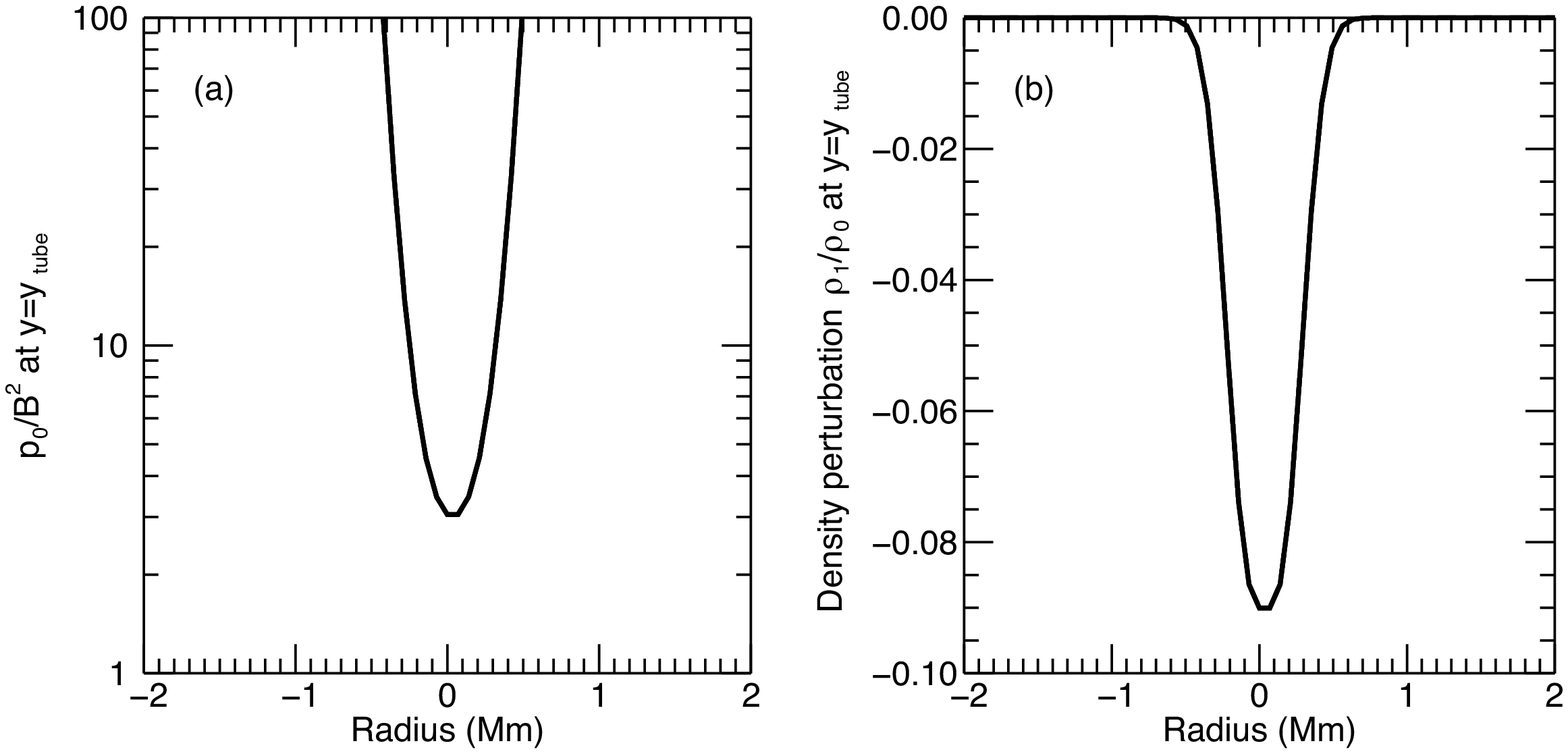}
\caption{Panel (a): Initial $p_{0}/B^{2}$ at $y=y_{tube}$ for the initial magnetic flux tube embedded in the convection zone for the three small scale simulations, Simulation 1 (the fully ionized simulation), Simulation 2 (the partially ionized simulation without ionization effects in the equation of state), and  Simulation 3 (the complete partially ionized model).  Panel (b): Initial density perturbation $\rho_{1}/\rho_{0}$, as in Equation (\ref{rho1}), in the tube for the same three simulations.
\label{fig:CZ_0.pdf}}
\end{center}
\end{figure}

To investigate the effects of partial ionization, we perform three separate simulations. Simulation 1 is the fully ionized plasma model (with the neutral reduced mass limit of $\mu_{m}=m_{i}$), as described in the previous section. 
To relate this work to the work of \citet{2006A&A...450..805L} we also include the model used in that paper, and call it Simulation 2. In Simulation 2, the ionization is calculated as a function of $\rho$ and $T$ using the Saha equation (\ref{eqn:saha}), but the second term in Equation (\ref{eqn:eos}) is ignored, so that ionization effects are not included in the energy calculation. In addition, this means that the convection zone profile used in \citet{2006A&A...450..805L}  does not include the variation of ionization in the equation for the adiabatic temperature gradient: it uses Equation (\ref{eqn:ad_temp_FIP}) rather than Equation (\ref{eqn:dta3}).  The initial background stratification for Simulation 2 is identical to Simulation 1, and so is not included in Figure \ref{fig:IC}. Simulation 3, as described in the previous section, uses the  partially ionized model with the full equation of state, and includes ionization and recombination in the  equation for the adiabatic temperature gradient.  As shown in Figure \ref{fig:IC}, Simulation 3 has a higher gas pressure at -1.8 Mm, where the flux tube is located, compared to Simulation 1.  So for the same field strength, the flux tube will have a higher plasma $\beta$ in the partially ionized simulation (3). We therefore vary the magnetic field so that the plasma $\beta$ has the same value at the center of the tube in all three simulations.  The rise speed of a buoyant flux tube scales as  $1/\sqrt{\beta}$ \citep{1996ApJ...464..999L}, and so by matching $\beta$ across simulations we ensure that the rise speeds are initially the same. For Simulations 1 and 2 we use $B_{0}=6000 ~ \textrm{G}$. For Simulation 3 we use $B_{0}=10900 ~ \textrm{G}$. The plasma $\beta$, calculated as $\beta=\mu_{0}p_{0}(y_{tube})/B^{2}(x,y_{tube})$, for the initial conditions of these three models is shown in Figure \ref{fig:CZ_0.pdf}, Panel (a). 

In all these simulations we add a perturbation $P_{1}(r)$ to  the background plasma pressure $P_{0}(y)$ such that 
$ \left(\nabla P_{1}\right)_{r} = \left(\mathbf{j}\wedge\mathbf{B}\right)_{r}$,
so that the tube is in radial force balance. Assuming that 
the flux tube is in thermal equilibrium with its surroundings  makes the tube less dense than the surrounding plasma and initiates its buoyant rise to the surface.  With this ``isothermal" assumption, the density perturbation, $\rho_{1}(r)$, which is added to the background density $
\rho_{0}(y)$, is given by 
\be
\rho_{1}/\rho_{0} = P_{1}/P_{0} \sim  -1/{2\beta}.
\label{rho1}
\ee
 This perturbation for the three simulations is shown in Figure \ref{fig:CZ_0.pdf}, Panel (b). 

\begin{table}

\begin{center}
  \begin{tabular}{| l | c | c | c | c | c | c | c | r | }
    \hline
    \hline
    Name & Model &  Reduced & Eq. of state & CZ profile & Tube radius &  $B_{0} (kG)$ & $\beta_{tube}$ & Flux\\ \
    & & mass $\mu_{m}$ & Eq.(\ref{eqn:eos})& $\left(\frac{\partial T}{\partial y}\right)_{a}$& & & & $\times10^{18}$ Mx\\
    \hline
     0 & FIP & $m_{f}m_{p}/2$ & - & Eq.(\ref{eqn:ad_temp_FIP})& - &  - &  - & \\
     1 & FIP & $m_{f}m_{p}$ &  -  & Eq.(\ref{eqn:ad_temp_FIP})& 0.3 Mm & 6 &   2&  5.4\\
     2 & PIP & Eq.(\ref{eqn:mu}) & No 2nd term & Eq.(\ref{eqn:ad_temp_FIP})& 0.3 Mm & 6& 2& 5.4\\
     3 & PIP & Eq. \ref{eqn:mu}) & 2nd term & Eq.(\ref{eqn:dta3})& 0.3 Mm &  10.9  & 2& 9.8\\
     \hline
     4 & FIP & $m_{f}m_{p} $ & - & Eq.(\ref{eqn:ad_temp_FIP})& 1.5 Mm & 15  & 4& 340.\\
     5 & PIP & Eq.(\ref{eqn:mu})  &  2nd term & Eq.(\ref{eqn:dta3}) & 1.5 Mm &  61.2  & 4& 1380.\\
     6 & FIP &   $m_{f}m_{p} $ &  - & Eq.(\ref{eqn:ad_temp_FIP})& 1.5 Mm &  0.49 & 40 & 11.\\
     7 & PIP & Eq.(\ref{eqn:mu})  &  2nd term & Eq.(\ref{eqn:dta3})&  1.5 Mm &  19.2  & 40 & 430.\\
       \hline
  \end{tabular}
  \caption{Simulations used in this paper. Simulation 0 is used only to highlight the choice of the reduced mass $\mu_{m}$ and is not run beyond $t=0 ~ \textrm{s}$. Simulation 2 has the same background stratification as the model used in \citet{2006A&A...450..805L}. FIP stands for `fully ionized plasma', and PIP for `partially ionized plasma' .
  }
\end{center}
\end{table}

\subsection{Emergence from the convection zone into the atmosphere}

In the three simulations the flux tubes rise to the surface due to their initial buoyancy and undergo a degree of horizontal expansion as they meet the convectively stable photosphere.  A contact layer is created when the tube's field meets the photosphere. As more field builds up at the surface, the layer becomes unstable to the magnetic Rayleigh Taylor instability \citep{1979SoPh...62...23A}. The criterion for this instability can be written as
\begin{equation}
-H_{p}\frac{\partial}{\partial y}ln(B) > H_{p}^{2} k^{2}\left(1+\frac{{l}^{2}}{{n}^{2}}\right)
-\frac{\gamma}{2}\beta\delta 
\label{eqn:MBI2}
\end{equation} 
\citep{2004A&A...426.1047A} where $H_{p}$ is the local pressure scale height and $\delta$ is the super-adiabatic 
excess \citep{STIX}, which is the  double logarithmic temperature gradient ($\partial ln T /\partial ln P$) minus its 
adiabatic value. In these simulations, $\delta$ is zero in the model convection zone, but negative in the model  photosphere. The wavenumber $k$ is the wavenumber parallel to the magnetic field in the horizontal plane, $l$ is the wavenumber perpendicular to the field in the horizontal plane, and $n$ is the wavenumber in the vertical direction. For the contact layer that is created, the left hand side in Equation (\ref{eqn:MBI2}) is positive, and acts to destabilize the layer. The last term on the right hand side is also positive and acts to stabilize the layer. 

\begin{figure}
\begin{center}
\includegraphics[width=0.8\textwidth]{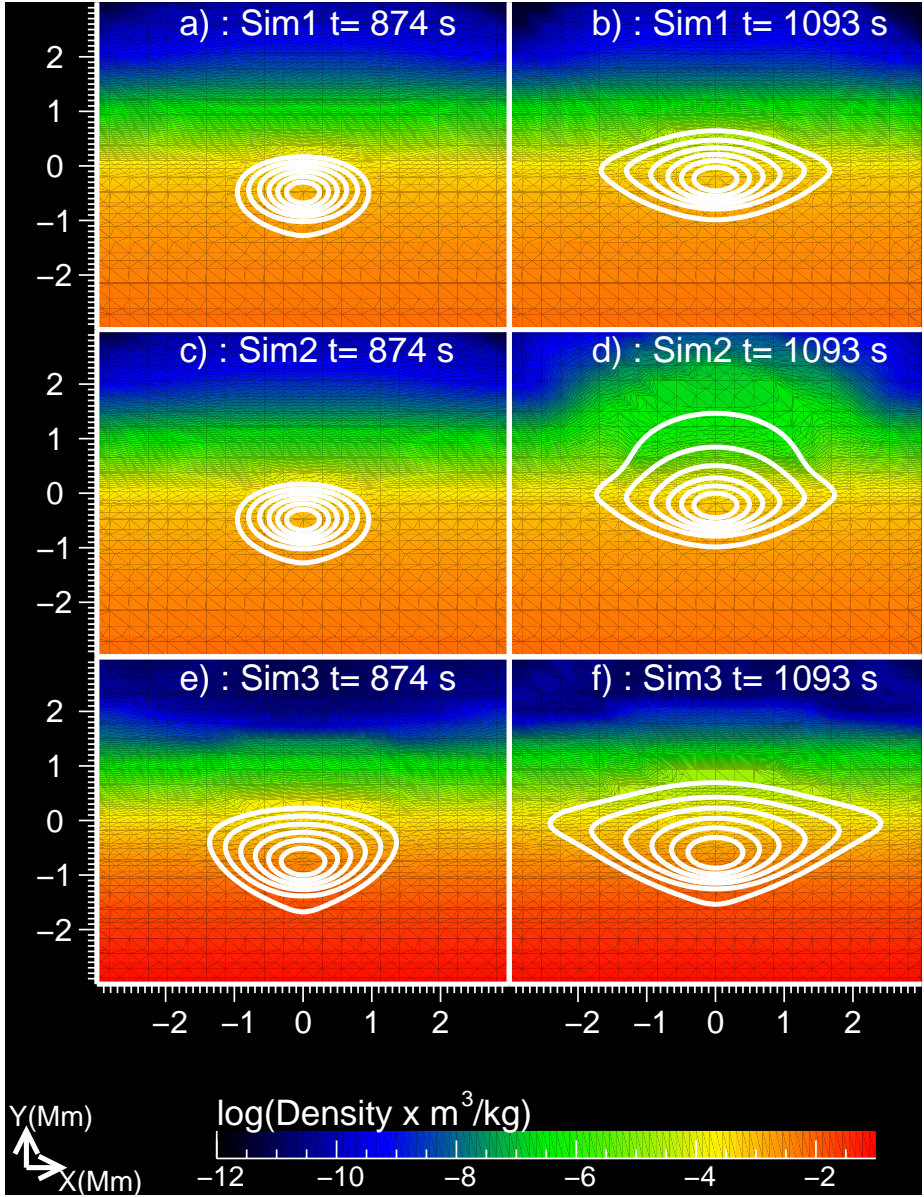}
\caption{Evolution of three small scale (ephemeral region size) flux tubes. The white lines show the in-plane field (constant contours of $A_{z}$) and the color contours show the log of density. The $A_{z}$ contours are at seven values regularly spaced between (and inclusive of) the
 extrema of $A_{z}$ and 0.1 of this extrema (the extrema contour is a single point located at the center of the tube). Top row:  Simulation 1, the fully ionized simulation. Middle row: Simulation 2, the partially ionized simulation without ionization effects in the equation of state. Bottom row: Simulation 3, the complete partially ionized model.
\label{fig:2D_rise_CZ.pdf}}
\end{center}
\end{figure}

Figure \ref{fig:2D_rise_CZ.pdf} shows the rise of the flux tubes in the convection zone in the three simulations. The contour lines show fieldlines in the plane, given by constant intervals of $A_{z}$, where $A_{z}$ is defined as the vector potential for the 2D vector $(B_{x},B_{y})=\nabla \wedge A_{z}\hat{\mathbf{e}}_{z}$ with boundary condition $A_{z}(\infty)=0$. The outer fieldlines shown here are the contours of $A_{z}=0.1\min(A_{z})$ where $A_{z}$ is always negative. The background color contour shows $\log({\rho} \times \textrm{m}^{3}/\textrm{kg})$. The tubes have risen to the surface with comparable speeds, which is to be
expected as they started with the same plasma $\beta$. The outer fieldlines are a different shape in 
Simulation 3, the partially ionized model simulation, than in Simulations 1 and 2. This is due to an increased horizontal expansion at the top of the tube in Simulation 3 relative to Simulations 1 and 2. This vertical gradient in horizontal expansion is caused by changes in ionization, which are not taken account in  Simulations 1 and  2. 

\begin{figure}
\begin{center}
\includegraphics[width=\textwidth]{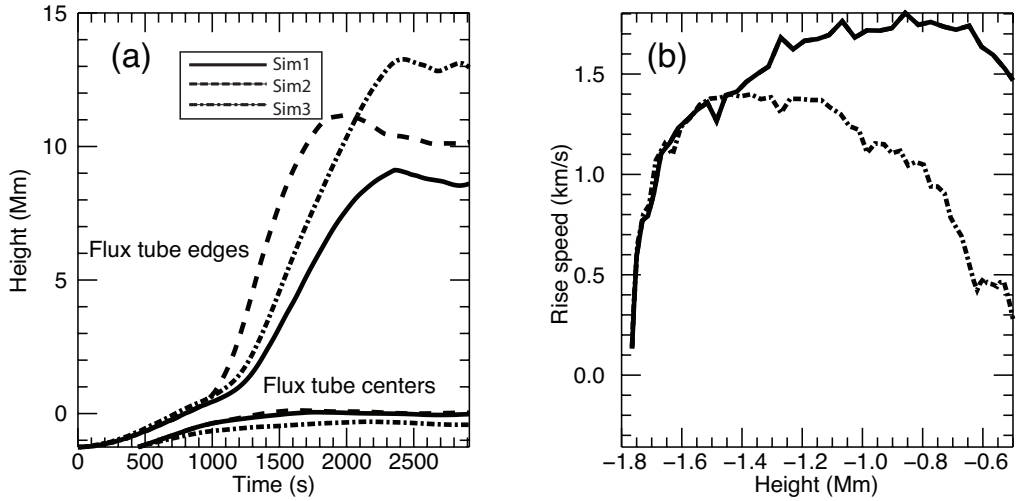}
\vspace{-70mm}
\caption{Panel (a): Height of the center and edge of the flux tubes as a function of time. The center is defined as the location of the minimum in the vector potential $A_{z}$. The edge is defined as the intersection of $A_{z}=0.1\min\left(A_{z}\right)$ with the $x=0$ line. Panel (b): Rise speed in the convection zone of the centers of the flux tubes for the same three simulations. The curves for Simulations 1 and 2 lie on top of each other in panel (b).
\label{fig:heights_SS.pdf}}
\end{center}
\end{figure}

Figure \ref{fig:heights_SS.pdf} shows the height of the center and edge of the tube in all three simulations. The center of the flux tube is defined as the location of the local extrema of $A_{z}$, as $A_{z}$ is initially a negative monotonically increasing function of radius for our twisted flux tubes. The edge of the flux tube is defined as the intersection of $A_{z}=0.1\min\left(A_{z}\right)$ with the $x=0$ line. In Simulation 3 (dot-dashed line), the flux tube's center  stops 0.5 Mm below the surface. On the other hand, in the fully ionized simulation (1) and the partially ionized simulation (2), the tube's center rises to $y=0$ (the surface). Figure \ref{fig:heights_SS.pdf}, Panel (b) shows the rise speed in the convection zone in all three simulations. All three simulations show an identical rise speed initially, as they have the same plasma $\beta$. The rise speed in Simulation 3 slows down relative to that of Simulations 1 and 2 due to changes in ionization during the rise.

\begin{figure}
\begin{center}
\includegraphics[width=\textwidth]{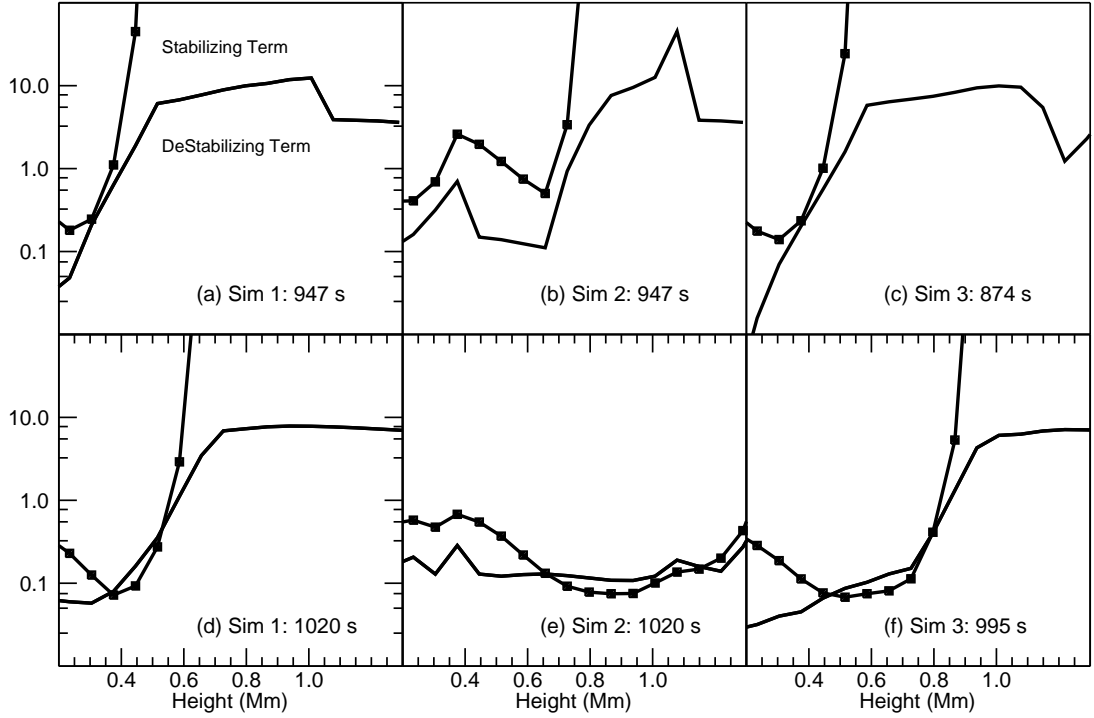}
\caption{Line plots along $x=0$ for the stabilizing ($-\frac{\gamma}{2}\beta\delta $) and destabilizing ($-H_{p}\frac{\partial}{\partial y}ln(B)$) terms  in the magnetic buoyancy instability inequality (\ref{eqn:MBI2}).  Left:  Simulation 1, the fully ionized simulation. Center: Simulation 2, the partially ionized simulation without ionization effects in the equation of state. Right: Simulation 3, the complete partially ionized model.
\label{fig:instabilities.pdf}}
\end{center}
\end{figure}

At around 1000-1020 s, all three simulations show the onset of the magnetic buoyancy instability. Figure \ref{fig:instabilities.pdf} shows the stabilizing ($-\frac{\gamma}{2}\beta\delta$) and the destabilizing ($-H_{p}\frac{\partial}{\partial y}ln(B)$) terms in Equation (\ref{eqn:MBI2}) for the three simulation at two different times. 
As magnetic field reaches the photosphere, the local plasma-$\beta$ decreases, which reduces the stabilizing term $
-\frac{\gamma}{2}\beta\delta$, until it becomes less than the destabilizing term $-H_{p}\frac{\partial}{\partial y}ln(B)$. At this point, the instability starts to develop, and magnetic field emerges into the atmosphere.
Simulation 2 shows a much smaller gradient in magnetic field than the other two simulations, indicated by a lower magnitude of the destabilizing term in Figure \ref{fig:instabilities.pdf}, Panel (e). In \cite{2006A&A...450..805L} we showed that the magnetic field was able to slip through the plasma due to the Pedersen resistivity, which dissipated currents perpendicular to the field.
In Simulation 2 we see the effect of this as a reduction in the vertical gradient of magnetic field in a layer above a height of 0.6 Mm above the surface. We do not see this in Simulation 1 as the model does not include Pedersen resistivity in that simulation. At this point in time, we do not see this in Simulation 3, as the rise of the tube in the convection zone in Simulation 3 was slowed due to the  ionization/recombination effects, but this reduction in the gradient in magnetic field is seen later in the simulation.

\begin{figure}
\begin{center}
\includegraphics[width=0.8\textwidth]{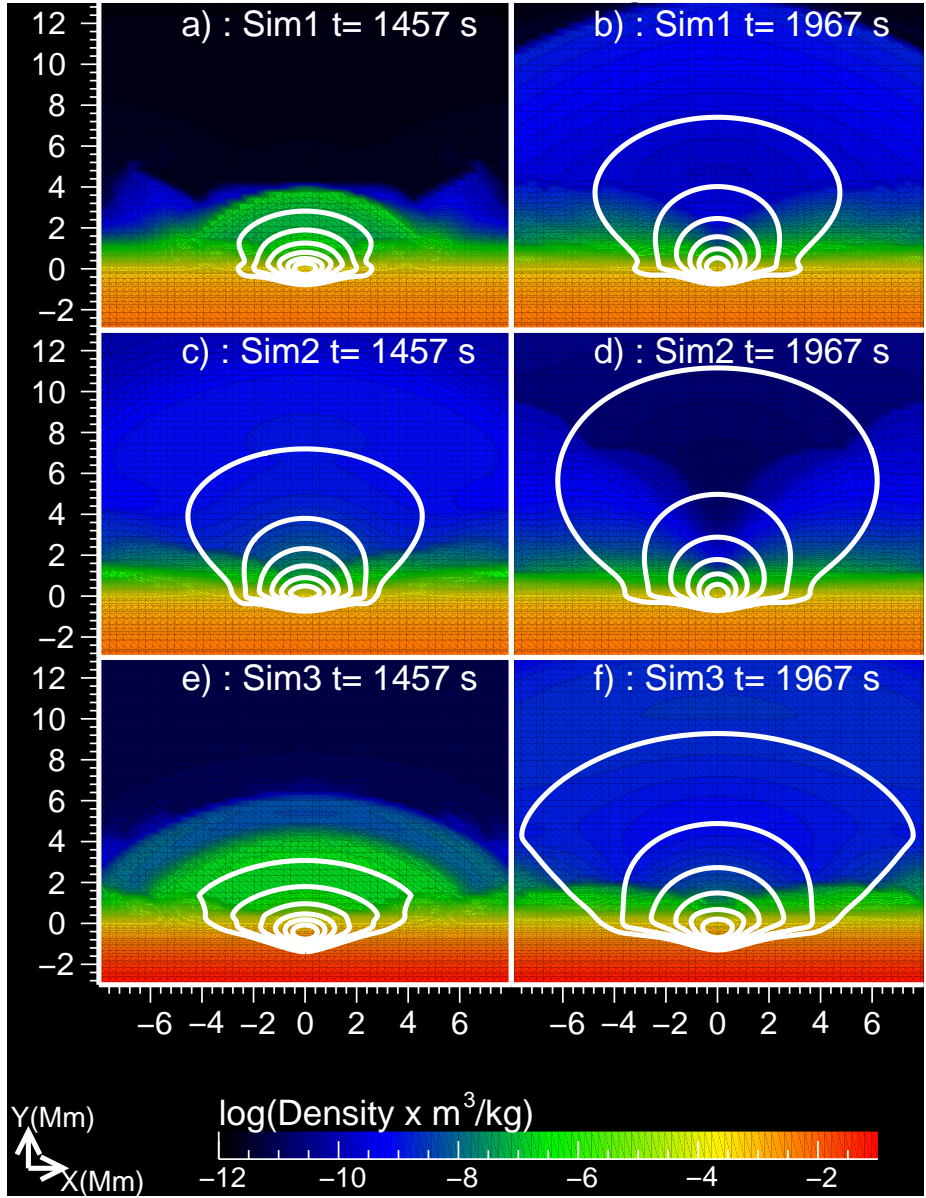}
\caption{Same as Figure \ref{fig:2D_rise_CZ.pdf} but at later times of 1457 s and 1967 s, showing later stage emergence of the flux tube in Simulations 1, 2 and 3.
\label{fig:2D_rise_atm.pdf}}
\end{center}
\end{figure}

\begin{figure}
\begin{center}
\vspace{-0mm}
\includegraphics[width=\textwidth]{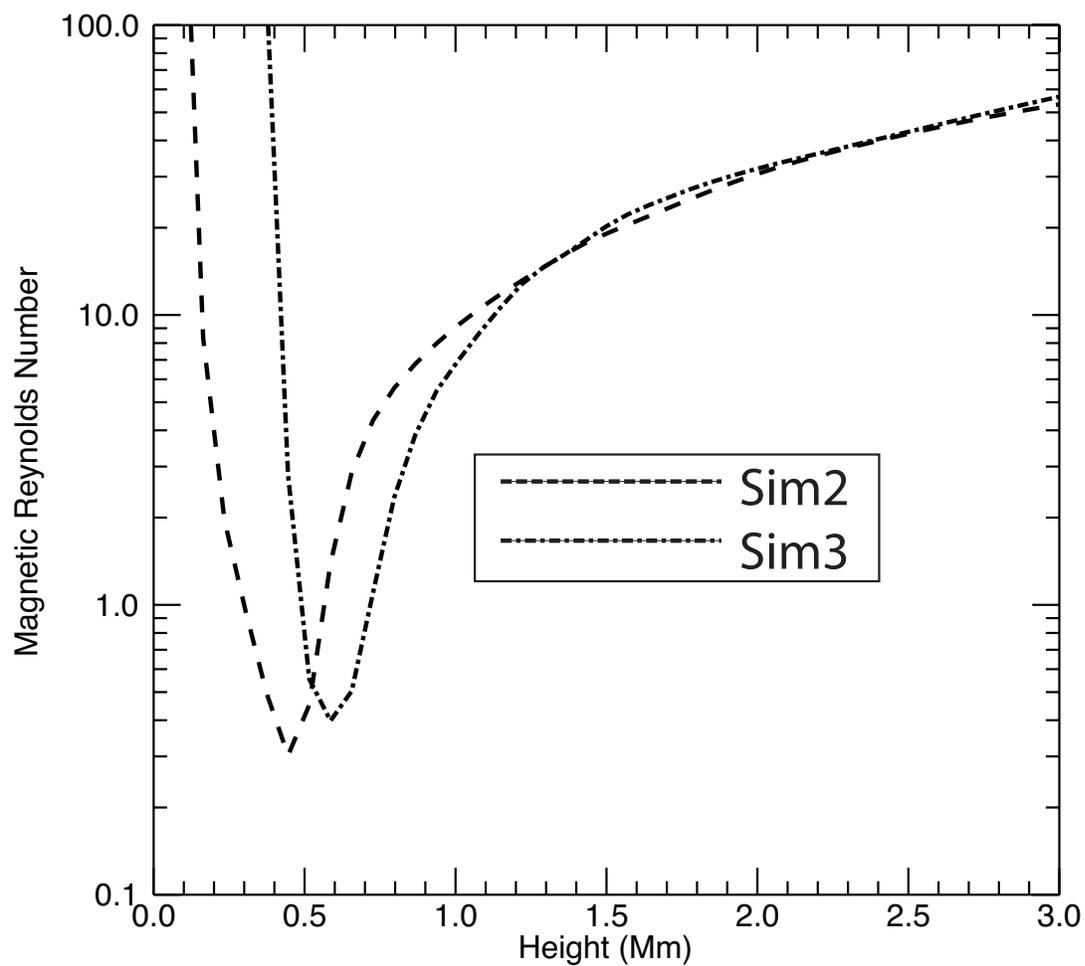}
\vspace{-10mm}
\caption{Magnetic Reynolds number (using the Pedersen resistivity) as a function of height at $x=0$ for the two partially ionized simulations.  The plot is taken at time 1359 s for Simulation 2, and at time 1724 s for Simulation 3.
\label{fig:reynolds_ohmic.pdf}}
\end{center}
\end{figure}

Figure \ref{fig:2D_rise_atm.pdf} shows the expansion of the magnetic field into the corona due to the magnetized Rayleigh Taylor instability, and Figure \ref{fig:reynolds_ohmic.pdf} shows the magnetic Reynolds number 
\be
R_{m} \sim \frac{|\mathbf{v}\wedge\mathbf{B}|}{|\eta_{p}\mathbf{j}_{\bot}|}
\ee
for the two partially ionized simulations (at time 1359 s for Simulation 2, and at time 1724 s for Simulation 3). The fully ionized simulation has a theoretically infinite Reynold's number (due to $\eta_{p}=0)$. At around 0.3 to 0.8 Mm above the surface, the two partially ionized simulations show a value of $R_{m}<1$. This 
means that the field's motion is dominated by diffusion rather than advection, and the dissipation increases the emergence of the field into the atmosphere. This result is in agreement with the previous simulations in \cite{2006A&A...450..805L}, even though in Simulation 3 we have a different equation for the specific energy density $\epsilon$ which allows for changes in ionization.

\begin{figure}
\begin{center}
\includegraphics[width=0.8\textwidth]{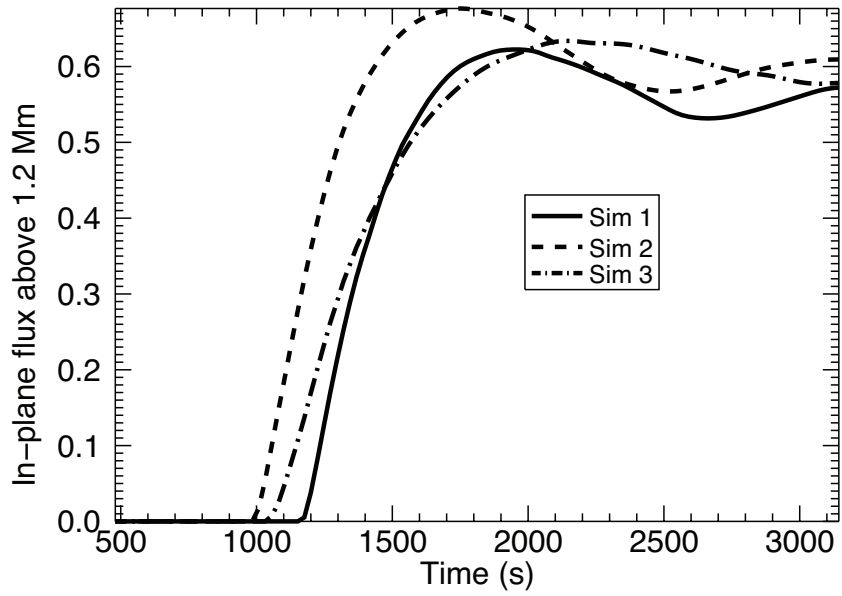}
\caption{In-plane flux above 1.2 Mm as a function of time, normalized to the in-plane flux in the initial flux tube, for Simulations 1, 2 and 3.
\label{fig:SS_flux}}
\end{center}
\end{figure}

As shown in Figure \ref{fig:heights_SS.pdf}, the upper edge of the tube expands further and faster into the corona in Simulations 2 and 3, as the increased dissipation due to ion-neutral collisions allows the field to diffuse through the plasma without having to lift material up. In the fully ionized simulation, the field must lift more plasma up with it, which 
slows its rise. 
Figure \ref{fig:SS_flux} shows the amount of in-plane flux that has emerged into the corona for each simulation, normalized to the initial amount of in-plane flux in each flux tube. This in-plane flux is calculated by
\be
\Phi(t)=[max(A_{z}(x,y))-min(A_{z}(x,y))]_{y>y_{1}}
\ee
where $y_{1}=1.2 ~ \textrm{Mm}$.
The flux $\Phi(t)$ is normalized to $|min(A_{z})|$ at t=0, so that it represents the fraction of initial in-plane flux that emerges above the height 1.2 Mm. All three simulations emerge above 60\% of the initial in-plane flux above this height. Due to the Pedersen dissipation, the two partially ionized simulations have peaks in the in-plane flux above 1.2 Mm higher than Simulation 1, the fully ionized simulation.

\begin{figure}
\begin{center}
\includegraphics[width=0.75\textwidth]{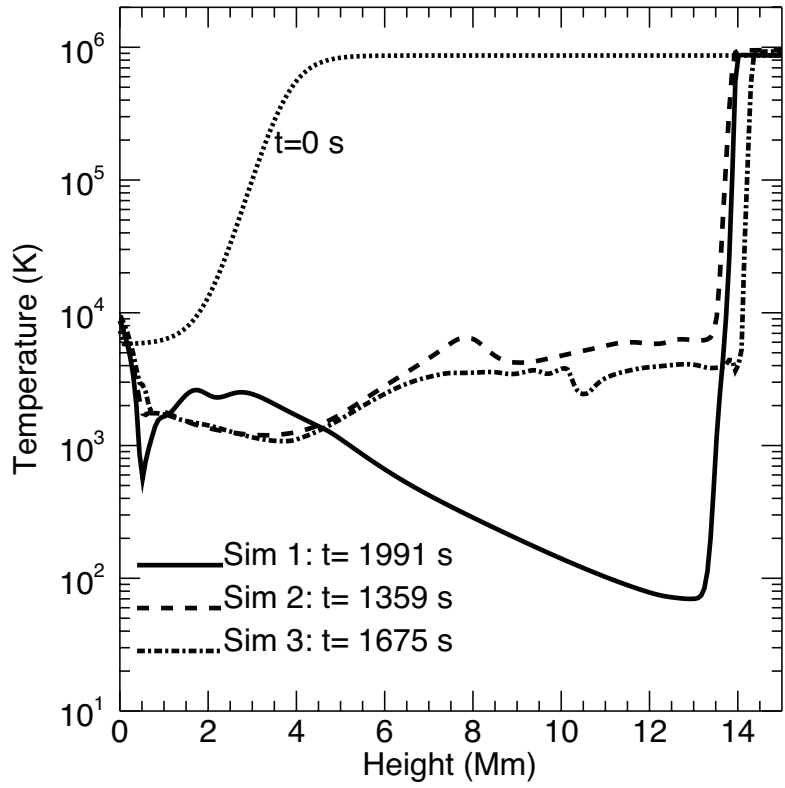}
\caption{Temperature as a function of height along the line $x=0$.  The times in each simulation are chosen so that the transition regions in each simulation are approximately co-spatial in height.
\label{fig:temp.pdf}}
\end{center}
\end{figure}

In the fully ionized model, the expansion of the plasma due to the emerging field is associated with a significant amount of cooling, and the plasma temperature drops to around 100 K within the emerging flux region, which can be seen in Figure \ref{fig:temp.pdf}. The transition region is pushed outward from 2 Mm to 14 Mm. The ion-neutral collision effects in the two partially ionized models (Simulations 2 and 3) are able to counteract this expansive cooling, and so this amount of cooling is not seen in the two partially ionized simulations, although some cooling to 1000 K is still seen and the transition region is still pushed outwards.  There are two effects which contribute to this result. The first is collisional ion-neutral heating. The second effect is the slippage of mass through the field,  due to the Pedersen resistivity, which allows the field to expand without expanding and cooling the plasma. These two effects ensure photospheric/chromospheric temperatures stay above 1000 K in the partially ionized simulations, as originally shown in \cite{2006A&A...450..805L}.
This result gives further evidence that ion-neutral collisions are an important source of heating in emerging active regions. 
Radiative-MHD numerical investigations  by \citet{2011A&A...530A.124L} show that, in 2D simulations, unrealistic cooling occurs which can only be counteracted by Joule heating. We have shown here that the additional Joule heating due to ion-neutral collisions is one way to achieve this necessary effect.
 
\begin{figure}
\begin{center}
\includegraphics[width=\textwidth]{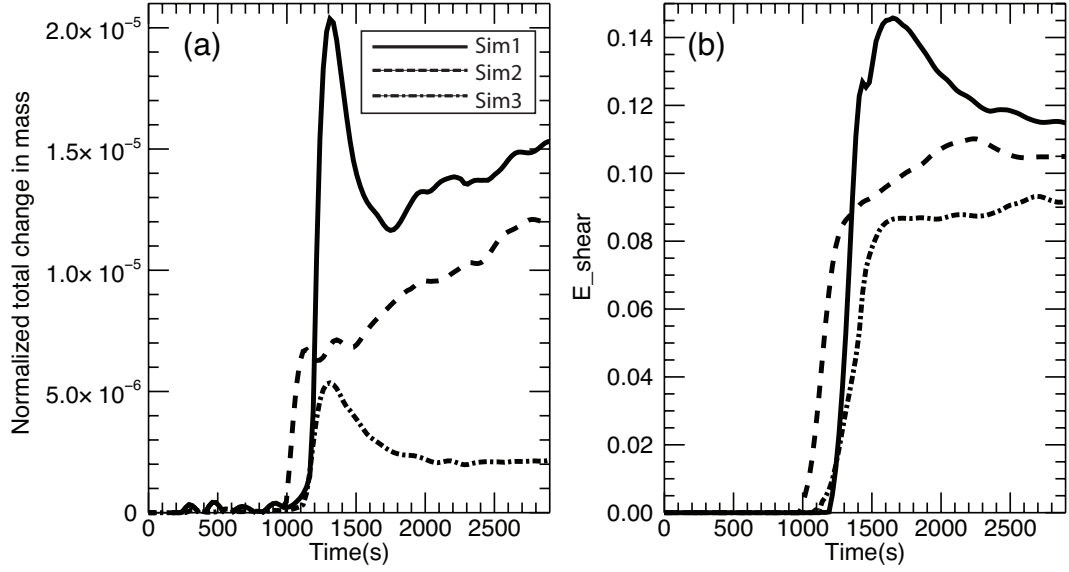}
\caption{Panel (a): Change in mass above 1.2 Mm, normalized to the total mass in the initial flux tube, as function of time. Panel (b):  $E_{shear}$, from Equation (\ref{eqn:E_shear}) for the same three simulations.
\label{fig:mass_shear.pdf}}
\end{center}
\end{figure}

We now investigate the effect of the increased dissipation due to ion-neutral collisions on the amount of mass and free energy supplied to the corona during flux emergence. In Simulation 3,  the density is higher at the initial tube location, as shown in Figure \ref{fig:IC}, and hence the flux tube contains more plasma than in the other two simulations (1 and 2).
We therefore calculate the change in mass in the corona, above a certain height during the flux emergence, normalized to the amount of mass initially in the flux tube. This diagnostic effectively estimates the fraction of the flux tube mass which is supplied to the corona. Figure \ref{fig:mass_shear.pdf} shows this diagnostic above 1.2 Mm in the three simulations. This shows that at t=2900 s, the fully ionized simulation (1) initially has lifted up approximately 7.7 times more of the tube's mass above 1.2 Mm than Simulation 3, but only 28\% more than Simulation 2. 

\begin{figure}
\begin{center}
\includegraphics[width=\textwidth]{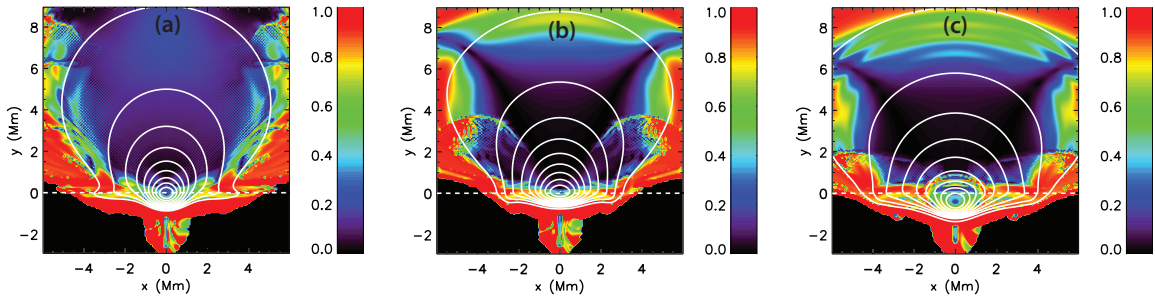}
\caption{The Lorentz force normalized by current density and magnetic field magnitudes $\frac{|\mathbf{j}\wedge\mathbf{B}|}{|\mathbf{j}||\mathbf{B}|}$, for Simulation 1 at 2377 s (panel a), Simulation 2 at 1560 s (panel b), and Simulation 3 at 1920 s (panel c). The times are shown so that in each figure the  $0.1 \min{(A_{z})}$ contour has reached approximately the same height. \label{fig:Lorentz}}
\end{center}
\end{figure}

Figure \ref{fig:Lorentz} shows the quantity $\frac{|\mathbf{j}\wedge\mathbf{B}|}{|\mathbf{j}||\mathbf{B}|}$, which is equivalent to $\sin(\theta)$, where $\theta$ is the angle between the current density and magnetic field, for the three simulations, during the post-emergence stage. A value of 0 means that all the currents are aligned with the field, i.e., the field is force-free.
This shows that the angle between the field and the current density is lower inside the expanding tube in Simulation 3 compared with Simulation 1. The Pedersen dissipation acts on the cross-field currents, thus reducing $\frac{|\mathbf{j}\wedge\mathbf{B}|}{|\mathbf{j}||\mathbf{B}|}$. This reduction in the Lorentz forcing component of the current density in part explains why so much less mass is supplied to the corona in Simulation 3, relative to Simulation 1. Simulation 2 also shows lower values of $\frac{|\mathbf{j}\wedge\mathbf{B}|}{|\mathbf{j}||\mathbf{B}|}$ inside the emerging active region,  relative to Simulation 1, again, explaining in part the reduction of mass supplied to the corona.

It is worth noting here that the Pedersen resistivity in the partially ionized model only effectively dissipates perpendicular currents \textit{inside} the emerging flux structure. The edge of the flux tube is a current sheet, with a large perpendicular current. However, for this current sheet, where the magnetic field decreases to zero, the Pedersen resistivity falls to zero as the current increases, and so there is limited dissipation of the current sheet.
In fact \citet{2009ApJ...705.1183A} showed that for current sheets with no guide field, the Pedersen resistivity acts to sharpen currents, rather than dissipate them. 

There are two effects of the ion-neutral collisions working in parallel in the two partially ionized models during the emergence. Firstly, due to the Pedersen resistivity, the mass can `slip' through the emerging magnetic field, and so less mass is lifted per unit amount of concave up flux in the partially ionized model relative to the fully ionized model. Secondly, cross-field currents are reduced by the Pedersen resistivity, and so the magnetic field cannot lift up as much mass in the partially ionized models as in the fully ionized model. For the parameters chosen for these simulation, we find that the result of these two effects is the reduction of the amount of mass supplied to the corona in both partially ionized models (2 and 3), relative to the fully ionized simulation (1).

Figure \ref{fig:mass_shear.pdf} also shows the amount of energy in the shear component of the field ($B_{z}$) normalized to the total amount of energy in the field.
\be
E_{shear} = \frac{ \int_{x=-22.5 Mm}^{x=22.5 Mm}{ \int_{y_{1}}^{y=42 Mm}{ B_{z}^{2}dxdy}}}{ \int_{x=-22.5 Mm}^{x=22.5 Mm}{ \int_{y_{1}}^{y=42 Mm}{ |\mathbf{B}|^{2}dxdy}}}
\label{eqn:E_shear}
\ee
where $y_{1}=1.2$ Mm.
In the fully ionized simulation (Simulation 1 - solid line), the shear field ($B_{z}$) is coupled to the mass, and we see more shear field supplied to the corona than in the other two simulations. At t=2900 s, Simulation 1 has a value of $E_{shear}$ 10\% higher than Simulation 2, and 25\% higher than Simulation 3. We postulate that Simulation 1 can emerge more shear flux, even though this means emerging more mass, because it has more Lorentz force.

Comparing Simulations 2 and 3  shows that including ionization effects in the equation of state significantly affects the emergence of flux. 
There is more horizontal expansion during the emergence process, both in the convection zone and the corona when the full equation of state is used. By 2900 s, Simulation 2 has emerged 
6 times the normalized mass into the corona that Simulation 3 has. Also, Simulation 2 has emerged 15\% more shear (in terms of $E_{shear}$) into the corona than Simulation 3.  We also note that the conclusions made in \citet{2006A&A...450..805L}, which used the model of Simulation 2, are repeated here with Simulation 3, which uses a more self-consistent model for the partially ionized plasma, in that it includes instantaneous ionization/recombination in the equation of state. These results include the increase in the amount of in-plane flux injected into the corona  (Simulations 2 and 3 emerge more in-plane flux than Simulation 1), and the additional Joule heating which is vital for maintaining a chromospheric temperature above 1000 K. 

In summary, we have performed three simulations of an emerging flux tube with a single value of $\beta$. Simulation 1 was a fully ionized model. Simulation 2 was a partially ionized model without the full equation of state, and Simulation 3 was a partially ionized model with the full equation of state, and a correct treatment of the adiabatic convection zone. We have shown that if a fully ionized model is used, the neutral limit for $\mu_{m} = m_{i}$ should be used to get the correct density in the corona.

These results show that although the presence of neutral Hydrogen can reduce the amount of mass supplied to the corona, which in principle can increase the likelihood of eruption, the amount of shear supplied to the corona in the partially ionized simulations is actually less than in the fully ionized simulations. Hence the likelihood of eruption is less in the partially ionized simulations. In all three simulations, the amount of magnetic energy in the shear component of the field is less than 20\% of the total magnetic energy in the field, and thus the coronal field is non-eruptive, just as in the simulations of \citet{Leake10}. For these flux tubes, the `slippage' of field through the nearly neutral lower atmosphere does not positively increase the likelihood of eruption.

\section{EMERGENCE OF LARGE SCALE FLUX TUBES}

The results of the previous section showed that including the ionization in the equation of state of the partially ionized model gave quantitatively different results from the model without the ionization terms. We performed the comparison in order to put the results in context with the  previous simulations of \citet{2006A&A...450..805L}. We know that the model used in Simulation 3, which includes the ionization in the equation of state, is the more self-consistent model of the two partially ionized models, and so we drop the model used in Simulation 2, which does not include the ionization term in the equation of state. We study now the fully ionized model and the  partially ionized model in terms of the emergence of active region size magnetic flux tubes.

The axial flux in the tube in the previous section's simulations was less than $10^{19}$ Mx, which is much smaller than the $10^{21}$ Mx size of a typical active region sunspot \citep{Priest,Zirin}. We must reconcile our initial conditions with realistic solar values to better test the effect of partial ionization on emerging active regions.

\begin{figure}
\begin{center}
\includegraphics[width=\textwidth]{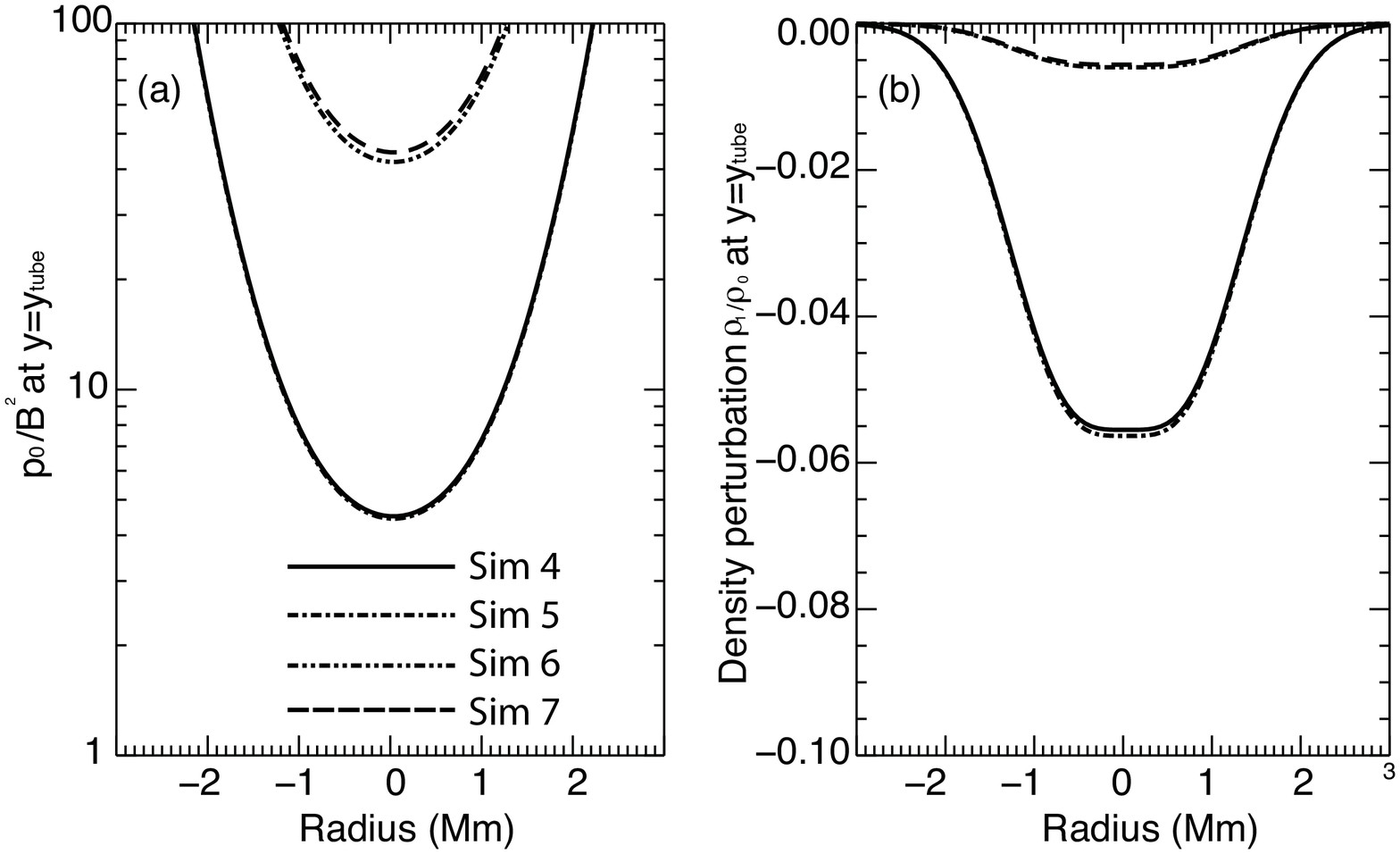}
\caption{Panel (a): Initial $p_{0}/B^{2}$ at $y=y_{tube}$  for the large scale flux tube simulations. Solid line is Simulation 4, the  $\beta=4$ fully ionized model. Dot-dash line is Simulation 5, the $\beta=4$ partially ionized model. Double-dot dash line is Simulation 6, the fully ionized model with $\beta \approx 40$. Long-dashed line is Simulation 7, the partially ionized model with $\beta \approx 40$. Panel (b): Initial density perturbation in the tube for the same three simulations.
\label{fig:initial_beta_LS.pdf}}
\end{center}
\end{figure}

To create an active-region scale flux tube, we increase the tube  radius to 1.5 Mm and place the tube at a depth of 
-4.95 Mm. The domain is increased to cover a range -56.25 to 56.25 Mm in $x$ and -15 to 97.5 Mm in $y$. The numerical resolution is kept the same as before at 70 km. The results of the previous section showed that flux tubes  with initially comparable $\beta$ in the flux tube start to rise with comparable speeds. 
In this section we run four  simulations, which are grouped into two pairs of comparable-$\beta$ simulations. The first pair consists of a fully ionized model (Simulation 4), and a partially ionized model (Simulation 5). Both simulations start with a flux tube which has $\beta=4$.
The second pair consists again of one fully ionized model (Simulation 6) and one partially ionized model (Simulation 7), both having a flux tube $\beta$ of 40. The simulations are briefly described in Table 1. The axial flux in the tube in  Simulation 5 is $1.4\times10^{21}$ Mx, and so is comparable with observed active region sunspots.
The two different values of $\beta$ of 4 and 40 are chosen to cover  a range of active region formation time. The simulations from the previous section show that flux tubes with an initial plasma $\beta$ of 2 emerge in less than 30 minutes, and so increasing the plasma $\beta$ in the tube to 40, we expect emergence on a timescale of approximately 2 hours, which is somewhat closer to the general formation time of active regions on the Sun (12-48 hours).  Figure \ref{fig:initial_beta_LS.pdf} shows the initial $\beta$ profiles and density perturbations, $\rho_{1}/\rho_{0}$, for these four simulations. 

\begin{figure}
\begin{center}
\includegraphics[width=0.8\textwidth]{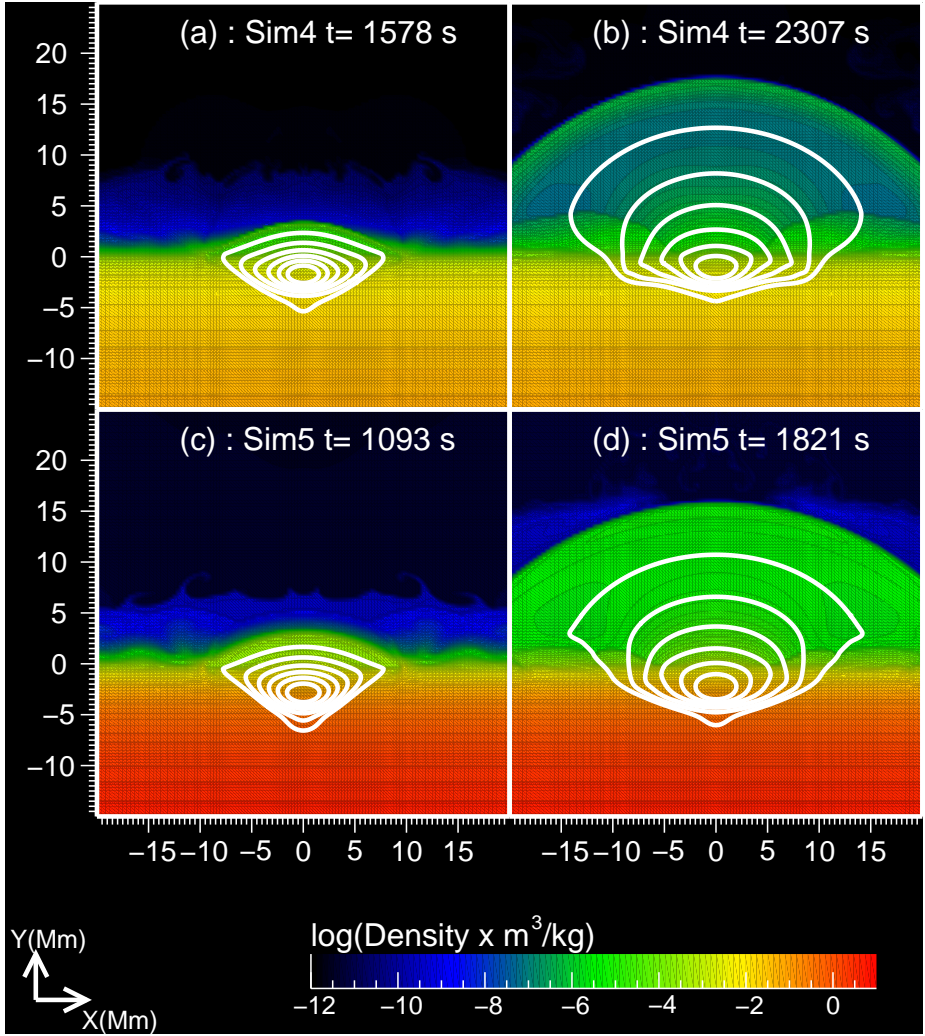}
\caption{Emergence of large scale (active region size) flux tubes, showing in-plane field (constant contours of $A_{z}$) and color contours of the log of density.   The $A_{z}$ contours are at seven values regularly spaced between (and inclusive of) the
 extrema of $A_{z}$ and 0.1 of this extrema (the extrema contour is a single point located at the center of the tube).  Top row:  Simulation 4, the fully ionized simulation with  $\beta=4$. Bottom row: Simulation 5, the partially ionized simulation with  $\beta=4$. 
\label{fig:2D_LS_a}}
\end{center}
\end{figure}

\begin{figure}
\begin{center}
\includegraphics[width=0.8\textwidth]{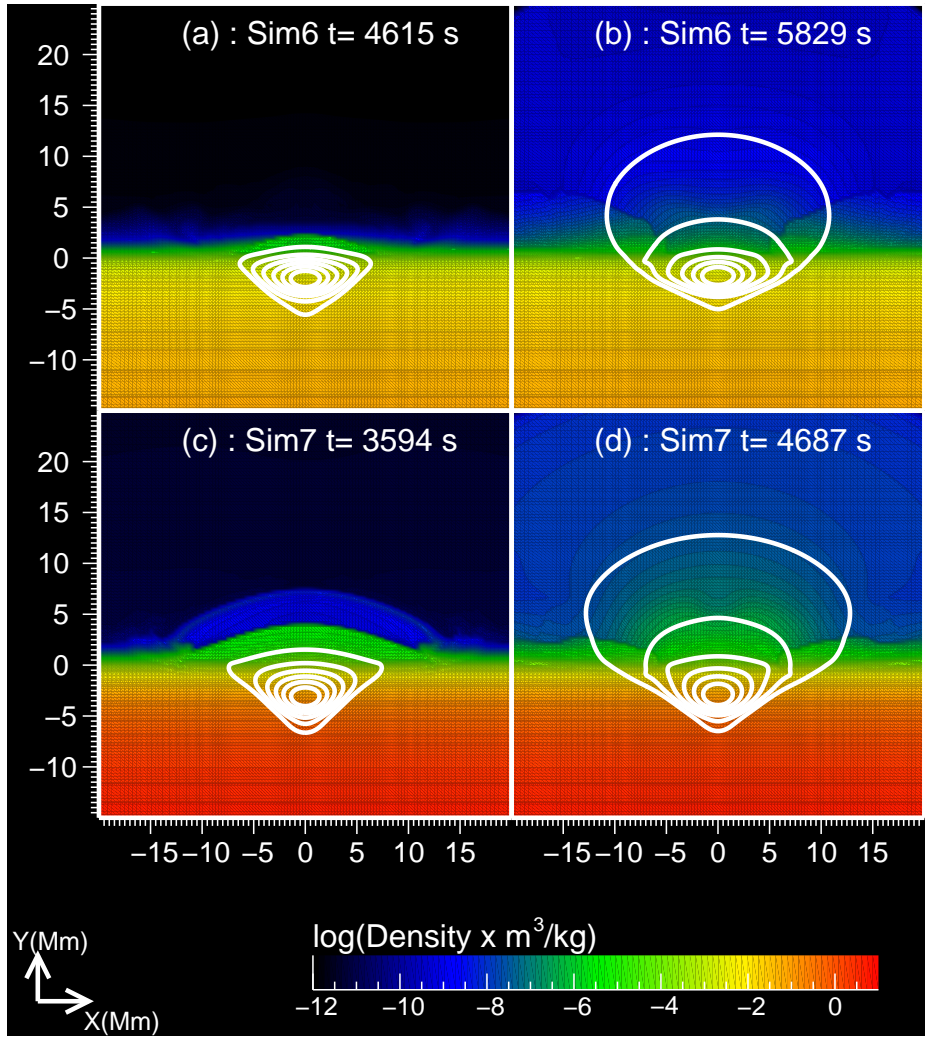}
\caption{Emergence of large scale (active region size) flux tubes, showing in-plane field (constant contours of $A_{z}$) and color contours of the log of density. Top row:  Simulation 6, the fully ionized simulation with $\beta=40$. Bottom row: Simulation 7, the partially ionized simulation $\beta=40$.
 \label{fig:2D_LS_b}}
\end{center}
\end{figure}

Figure \ref{fig:2D_LS_a} shows the in-plane magnetic field as regular contours of $A_{z}$, and log of density as a color contour, at two different times in Simulations 4 and 5 (the $\beta=4$ simulations). The first time (left panels) is when the tube has reached the surface, and built up enough magnetic field to trigger the magnetic buoyancy instability. The second time (right panels) is at a later stage, when the outer 10\% of the in-plane flux has reached approximately 10 Mm above the surface.  Figure \ref{fig:2D_LS_b} shows the same phases of the emergence process but for the two simulations 6 and 7, where $\beta=40$. The high $\beta$ in these simulations means that the initial rise-speed is slow compared to the low $\beta$ simulations 4 and 5, and it takes longer to build up enough field at the surface to trigger the instability which drives field into the corona. 

\begin{figure}
\begin{center}
\includegraphics[width=\textwidth]{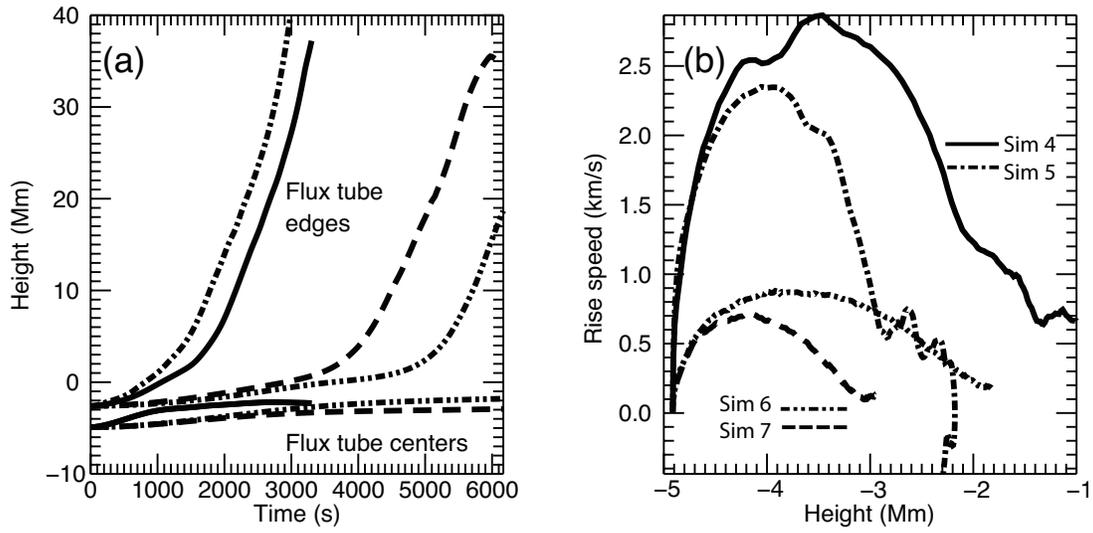}
\caption{Panel (a): Height of the center and edges of the flux tube as function of time. The curves for the tube centers in Simulations 4 and 5 lie on top of each other. Panel (b): Axial rise speeds in the convection zone for the same four simulations.
\label{fig:heights_LS.pdf}}
\end{center}
\end{figure}

Figure \ref{fig:heights_LS.pdf}, Panel (a) shows the height of the tube center and tube edge for the large scale simulations. As in the small scale simulations, the tube centers are confined to just below the surface, while the edges expand in to the field-free corona. 
Figure \ref{fig:heights_LS.pdf}, Panel (b) shows the rise speeds in the convection zone. 
Firstly, we note that these simulations confirm that the convection zone rise speed scales with $1/\sqrt{\beta}$ \citep{1996ApJ...464..999L}; Taking the two fully ionized simulations (4 and 6), and calculating the ratio of the maximum rise speeds in the convection zone gives a value of 3.3, which is approximately equal to the inverse of the ratio of $\sqrt{\beta}$ in those two simulations ($\sqrt{40/4}$). The same applies to the rise speeds in the two partially ionized simulations (5 and 7).

Comparing Simulation 4 to Simulation 5 (the two $\beta=4$ models, solid and dot-dash line respectively), the rise speed peaks at a value of 2.3 km/s in the partially ionized model (Simulation 5), and a value of 2.9 km/s in the fully ionized model (Simulation 4). This difference in rise speeds also occurred in the small scale simulations, and is a consequence of the ionization and recombination during the rise of the tube in the partially ionized convection zone. This result is repeated in the $\beta=40$ simulations (Simulation 6: double-dot-dash line,  and Simulation 7: long-dash line), with the fully ionized simulation obtaining 
a higher rise speed than the partially ionized simulation.  This may explain why the flux tube axis settles lower in the convection zone in Simulation 7 compared to Simulation 6. This effect may be important for the formation of coronal structures in 3D, as it is thought that the evolution of the flux tube axis is important for the formation of sheared structures in 3D.

\begin{figure}
\begin{center}
\includegraphics[width=0.8\textwidth]{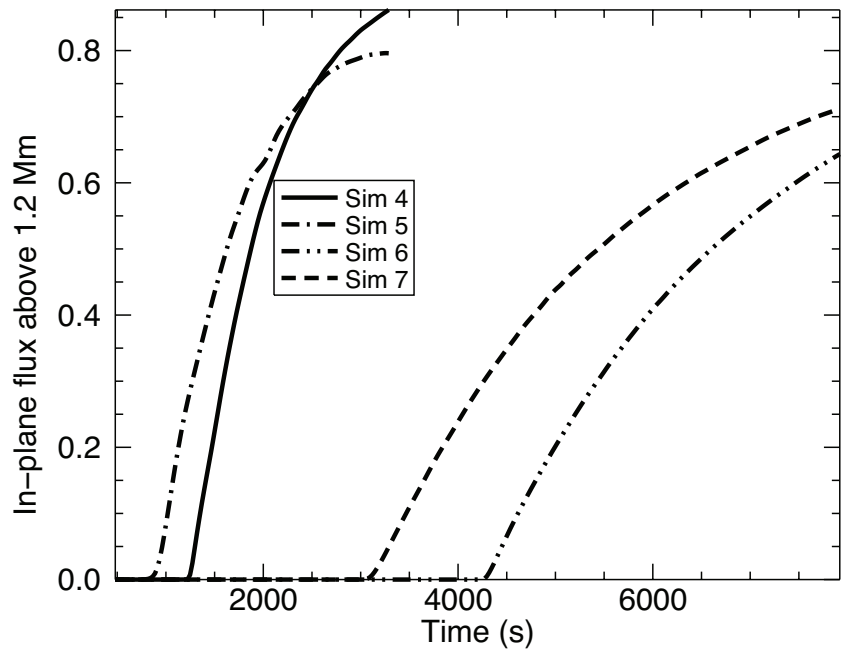}
\caption{The amount of in-plane flux above 1.2 Mm as a function of time, normalized to the integral of $A_{z}$ below $y=0$ at t = 0 s, for Simulations 4, 5, 6 and 7.
\label{fig:LS_flux}}
\end{center}
\end{figure}

The amount of normalized in-plane flux ($\Phi$) supplied to the corona above 1.2 Mm is shown in Figure \ref{fig:LS_flux}. 
Comparing $\Phi(t)$ for the low and high $\beta$ simulations, we see that more of the original in-plane flux emerges in the low $\beta$ simulations (4 and 5) than in the high $\beta$ simulations (6 and 7). Still there is almost 50\% of the original in-plane flux supplied to the corona in the high $\beta$ case,
which is comparable to the amount of flux which emerges above 1.2 Mm in the small-scale simulations of the previous section.
Comparing the fully ionized simulations to the partially ionized simulations (4 vs. 5 and 6 vs. 7 ) it is clear that the flux emerges earlier in the partially ionized simulations (5 and 7), compared to the fully ionized simulations (4 and 6).

\begin{figure}
\begin{center}
\includegraphics[width=0.75\textwidth]{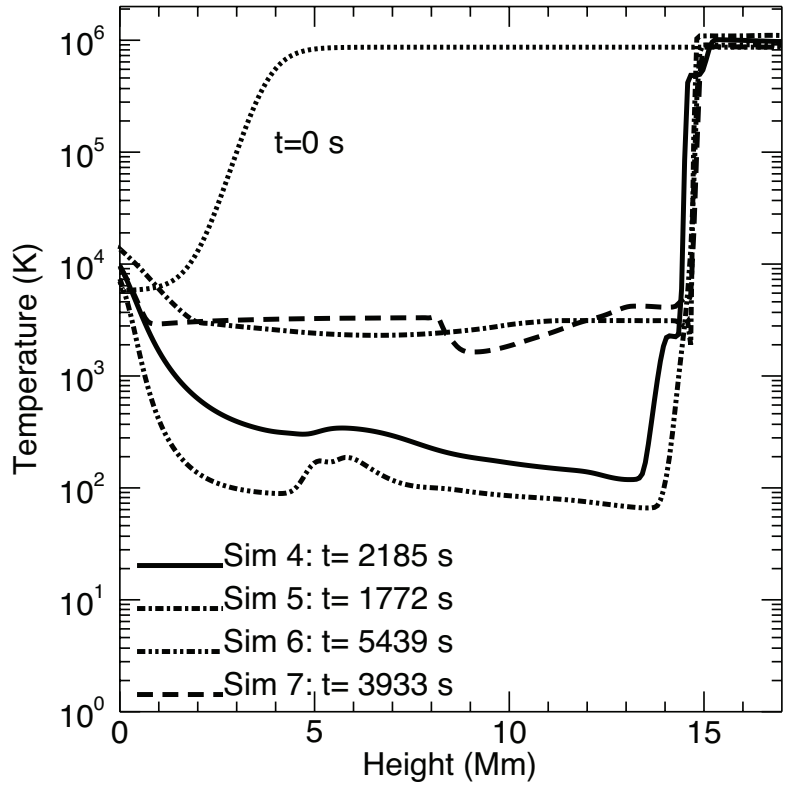}
\caption{Temperature as a function of height at $x=0$ for all four large scale simulations. 
The times in each simulation are chosen so that the transition regions in each simulation are approximately co-spatial in the y-direction.
\label{fig:joule_temp_LS.pdf}}
\end{center}
\end{figure}

The resulting temperature profiles are shown in Figure \ref{fig:joule_temp_LS.pdf}. As in the small scale simulations, the partially ionized simulations do not see the drastic cooling below 1000 K in the chromosphere that is seen in the fully ionized simulations. This is a result of both the collisional heating, and the dissipation by Pedersen resistivity reducing the expansion of the plasma as the field expands, a result which we found in the smaller-scale simulations of the previous section. We have therefore shown that this result applies to the larger active region and slower emergence of these large scale simulations (4 through 7).

\begin{figure}
\begin{center}
\includegraphics[width=\textwidth]{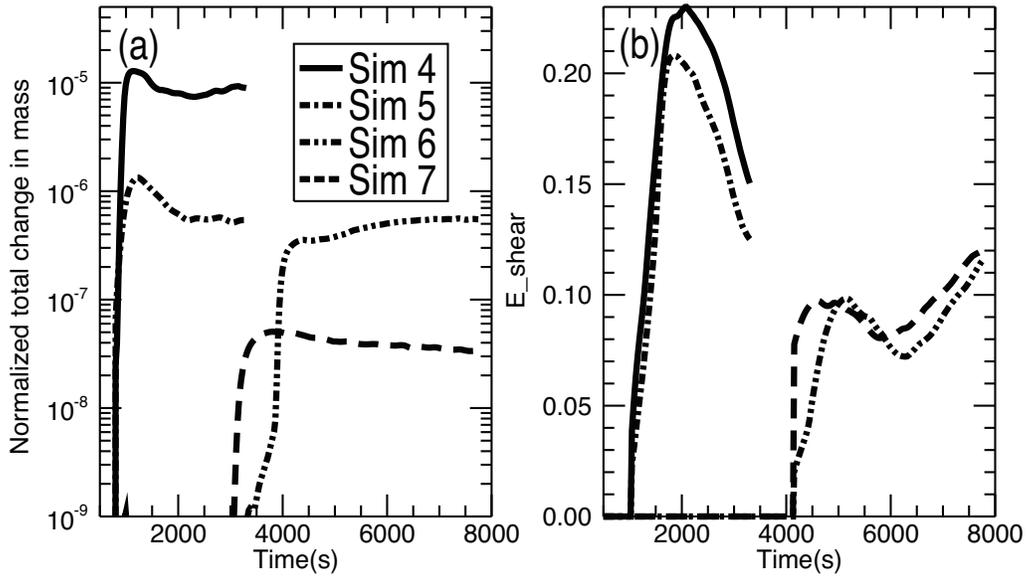}
\caption{Panel (a): Change in mass above 1.2 Mm, normalized to the total mass in the initial flux tube, as a function of time. Solid line: Simulation 4, the fully ionized simulation with $\beta=4$. Dot--dashed line: Simulation 5, the partially ionized simulation with the same $\beta$ as Simulation 4.  Double-dot-dashed line: Simulation 6, the fully ionized model with $\beta=40$. Long dashed line: Simulation 7, the partially ionized simulation with the same $\beta$ as Simulation 6. Panel (b):  $E_{shear}$, from Equation (\ref{eqn:E_shear}) for the same four simulations.
\label{fig:mass_shear_LS.pdf}}
\end{center}
\end{figure}

Figure \ref{fig:mass_shear_LS.pdf} shows the normalized change in mass above 1.2 Mm in the four large scale  simulations.  Comparing the two $\beta=4$ simulations (4 and 5), we find that the fully ionized simulation supplies approximately 10 times more of the flux tube mass to the corona than the partially ionized simulation, a result which is repeated for the two $\beta=40$ simulations (6 and 7). This, again is a direct consequence of the Pedersen dissipation in the partially ionized models.

It is worth comparing the amount of flux tube mass supplied to the corona in these large scale simulations to that in the small scale simulations. The two large scale $\beta=4$ simulations supply approximately the same amount of flux tube mass above 1.2 Mm ($10^{-5}$ for the fully ionized model and $10^{-6}$ for the partially ionized) as the small scale simulations, which have $\beta\approx 2$. The $\beta=40$ large scale simulations supply more than an order of magnitude less flux tube mass to the corona than the $\beta=4$ simulations ($5\times10^{-7}$ for the fully ionized and $4\times10^{-8}$ for the partially ionized model). So in general the amount of flux tube mass supplied above 1.2 Mm during flux emergence is inversely proportional to the initial $\beta$ in the flux tubes at t=0, and hence is dependent on the strength of the magnetic field in the tubes, as one would expect.

Also shown in Figure \ref{fig:mass_shear_LS.pdf} is $E_{shear}$ above 1.2 Mm. For the $\beta=4$ simulations, the fully ionized simulation supplies 15\% more free energy at t= 2900 s than the partially ionized simulation. There is no significant difference between the two $\beta=40$ simulations. In these large scale simulations, less than 25\% of the magnetic energy in the corona is in the shear component, which broadly agrees with the results of the fully ionized simulations of \citet{2010ApJ...722..550L}. Although the free energy does exceed 20\% of the magnetic energy in the $\beta=4$ simulations, it quickly falls below this value. The $\beta=40$ simulations stay below 20\% for the duration of the simulations. The resulting coronal configurations in these 2.5D simulations are non-eruptive, despite the decrease in mass supplied to the corona due to the presence of neutral material in the lower solar atmosphere.

The successful emergence of flux tubes with active region values of magnetic flux ($10^{21}$ Mx) has been achieved, and been shown to be qualitatively similar to the emergence of smaller scale flux tubes, which have been studied extensively. Furthermore, the rise speed of these large scale flux tubes is affected by the partially ionized region of the solar atmosphere, as is the amount of mass and shear supplied to the corona, even though the initial tube radius (1.5 Mm) is comparable to the size of the region where the Pedersen resistivity is important (from the surface to about 1 Mm above the surface).

\section{CONCLUSIONS}

We have investigated the effect of neutral Hydrogen in the lower solar atmosphere on the emergence of sub-surface magnetic flux into the solar atmosphere, using 2.5D MHD models  modified to include a variable ionization state, and the anisotropy in Ohm's law caused by ion-neutral collisions.
Previous simulations using a fully ionized MHD model \citep{2010ApJ...722..550L} have shown that the amount of free energy is insufficient to drive a CME when flux tubes emerge  in this 2.5D cartesian setup. The presence of too much mass and the lack of a strong enough shear field yielded no further expansion of the flux rope structure, which was ultimately constrained by overlying field. Typically the maximum amount of energy in the shear field was 10-15\% of the magnetic energy. The main aim of this paper was to investigate the effect of the partially ionized atmosphere on the likelihood of creating eruptive magnetic field from emerging convection zone flux tube.

The single-fluid MHD equations were modified to include the presence of neutral Hydrogen.  This led to a modified induction equation, where perpendicular currents were dissipated by ion-neutral collisions and parallel currents were dissipated by electron-ion and electron-neutral collisions. It also led to a source term  related to ionization state in the equation for the specific internal energy density. 
Previous simulations \citep{2005A&A...442.1091L,2006A&A...450..805L,2007ApJ...666..541A} with these effects used a simple equation of state which did not take into account ionization/recombination, and thus used a model for an adiabatic convection zone which did not include partial ionization. In this paper we included the change in ionization in our equation of state, and therefore a more realistic convection zone profile.

In the partially ionized simulations, the dissipation due to the Pedersen resistivity and the associated collisional heating was concentrated in the lower atmosphere, below the transition region. In the fully ionized simulations, the rapid expansion of the emerging field was associated with a rapid expansion and cooling of the plasma in the emerging region, which resulted in temperatures below 1000 K, which we consider to be too low for the real Sun. In the partially ionized simulations, we did not see such cooling below 1000 K. There are two effects which contribute to this. The first is the fact that as the field emerges into the corona in the partially ionized simulations, the Pedersen resistivity allows the field to `slip' through the plasma, and this reduces the amount of expansion and cooling seen in the emerging region. The second is that any cooling will be countered by the collision dissipation.
This result gives further evidence that ion-neutral collisions are an important source of heating in emerging active regions, as was suggest in the radiative-MHD simulations of \citet{2011A&A...530A.124L}. 

The result of the increased dissipation in the lower atmosphere in the partially ionized simulations led to a number of effects. Firstly, for the small scale simulations and the $\beta=40$ large scale simulations, the amount of in-plane flux supplied to the corona increased compared to the fully ionized models. The Pedersen dissipation in the partially ionized simulations  led  to a decrease in the Lorentz forcing component of the current density inside emerging active regions, and a reduction in the  mass lifted into the corona by the emerging flux. In the fully ionized simulations, there was more of the flux tube mass supplied to the corona. Coupled to this was an increase in shear field supplied to the corona in the fully ionized simulation. As a result, the likelihood of eruption was not increased by the effects of neutral Hydrogen in the lower atmosphere, but instead reduced.

We performed studies with both small scale (radius of 0.3 Mm and fluxes of $5-10\times10^{18}$ Mx), and large scale (radius of 1.5 Mm and fluxes of $10^{19}-10^{21}$ Mx) flux tubes. The larger flux tubes resulted in active regions of size 20 Mm, based on the separation of the two intersections of the 10\% flux contour with the surface $y=0$. For the smaller flux tubes the active region size was typically 10 Mm. Interestingly, increasing the initial flux tube radius by a factor of 5 only resulted in an increase in the resultant active region size by a factor of 2.

Results of our small scale flux emergence simulations showed that by including the ionization in the equation of state,  the rise speed was effectively reduced by ionization and recombination in the upper convection zone, leaving the axis of the tube lower down in the convection zone after the emergence. Also in the simulations where the ionization was included in the equation of state, the amount of mass and shear flux supplied to the corona was decreased relative to the partially ionized simulation which did not include ionization in the equation of state.

 For the larger scale simulations we used two different vales for $\beta$ in the flux tube. The timescale of emergence depended on the $\beta$ value. For $\beta=4$ the flux tube edges took 50 minutes to reach a height of 35 Mm, whereas for $\beta=40$ the emergence took 100 minutes to reach the same height. This delay was almost entirely accounted for by the slower rise speed in the convection zone, which scaled with $\sqrt{1/\beta}$. In all the low and high $\beta$ simulations, more than 55\% of the initial in-plane flux emerges into the corona. Even though the flux tube's radius in the large scale simulation was 10 times larger than the photospheric scale height of 150 km, the emergence was qualitatively similar to the emergence of smaller flux tubes. 
 
For these 2.5D simulations, the presence of neutral Hydrogen in the lower atmosphere, and the associated `slippage' of emerging magnetic field (or equivalently the associated motion of plasma across fieldlines) did not increase the amount of shear flux supplied to the corona, and therefore we conclude that the effects of neutral Hydrogen do not increase
 the likelihood of eruption. We also conclude that 3D plasma motions along the axis of the flux tubes are more likely to destabilize coronal magnetic field in emerging active regions. However, the  stability of flux ropes formed during flux emergence, and the subsequent likelihood of eruption, will be dependent on the currents associated with the emerging fields, and therefore should be strongly dependent on the nature of the current dissipation mechanism. We propose in future work to investigate both 3D effects and the effects of partial ionization.
Note that these conclusions address how the presence of neutral Hydrogen affects the supply of free energy, or sheared field, into the corona by the emergence of magnetic flux into a field-free corona. Hence the only possible source of free energy in these simulations is the newly emerging flux. In the case where a pre-existing sheared structure is already formed in the corona, the emergence of new flux can play the role of destabilizing the magnetic field and causing an eruption. 


\begin{acknowledgments}
\noindent{Acknowledgements:}
This work has been supported by the NASA Living With a Star,  Solar \& Heliospheric 
Physics programs,  the ONR 6.1 Program, and the NRL-Hinode analysis program. 
The simulations were performed under a grant of computer time from the DoD HPC program.
\end{acknowledgments}

\bibliography{main_bib2}

\eject

\clearpage

\end{document}